\documentclass{cmspaper}

\usepackage{amsmath,amssymb}
\usepackage{graphicx,graphics}
\usepackage{xspace}
\usepackage[bookmarksnumbered,bookmarksopen,bookmarksopenlevel=1,colorlinks=false,pdfborder=0,plainpages=false,pdfpagelabels]{hyperref}
\usepackage[numbers,sort&compress]{natbib}
\usepackage{hypernat}

\hypersetup{%
  pdfauthor={K.~Rabbertz},%
  pdftitle={QCD Physics Potential of CMS},%
  pdfsubject={CMS},%
  pdfkeywords={CMS, physics, QCD}
}

\begin{document}

%
% Set standard definitions etc.
%
% --- force emacs to latex mode -*-latex-*-
%%%%%%%%%%%%%%%%%%%%%%%%%%%%%%%%%%%%%%%%%%%%%%%%%%%%%%%%%%%%%%%%%%%%
%
%  Common definitions
%
%  N.B. use of \providecommand rather than \newcommand means
%       that a definition is ignored if already specified
%
%                                              L. Taylor 18 Feb 2005
%%%%%%%%%%%%%%%%%%%%%%%%%%%%%%%%%%%%%%%%%%%%%%%%%%%%%%%%%%%%%%%%%%%%

% Some shorthand
% turn off italics
\newcommand {\etal}{\mbox{et al.}\xspace} %et al. - no preceding comma
\newcommand {\ie}{\mbox{i.e.}\xspace}     %i.e.
\newcommand {\eg}{\mbox{e.g.}\xspace}     %e.g.
\newcommand {\etc}{\mbox{etc.}\xspace}     %etc.
\newcommand {\vs}{\mbox{\sl vs.}\xspace}      %vs.
\newcommand {\mdash}{\ensuremath{\mathrm{-}}} % for use within formulas

% some terms whose definition we may change
\newcommand {\Lone}{Level-1\xspace} % Level-1 or L1 ?
\newcommand {\Ltwo}{Level-2\xspace}
\newcommand {\Lthree}{Level-3\xspace}

% Some software programs (alphabetized)
\providecommand{\ACERMC} {\textsc{AcerMC}\xspace}
\providecommand{\ALPGEN} {{\textsc{alpgen}}\xspace}
\providecommand{\CHARYBDIS} {{\textsc{charybdis}}\xspace}
\providecommand{\CMKIN} {\textsc{cmkin}\xspace}
\providecommand{\CMSIM} {{\textsc{cmsim}}\xspace}
\providecommand{\CMSSW} {{\textsc{cmssw}}\xspace}
\providecommand{\COBRA} {{\textsc{cobra}}\xspace}
\providecommand{\COCOA} {{\textsc{cocoa}}\xspace}
\providecommand{\COMPHEP} {\textsc{CompHEP}\xspace}
\providecommand{\EVTGEN} {{\textsc{evtgen}}\xspace}
\providecommand{\FAMOS} {{\textsc{famos}}\xspace}
\providecommand{\GARCON} {\textsc{garcon}\xspace}
\providecommand{\GARFIELD} {{\textsc{garfield}}\xspace}
\providecommand{\GEANE} {{\textsc{geane}}\xspace}
\providecommand{\GEANTfour} {{\textsc{geant4}}\xspace}
\providecommand{\GEANTthree} {{\textsc{geant3}}\xspace}
\providecommand{\GEANT} {{\textsc{geant}}\xspace}
\providecommand{\HDECAY} {\textsc{hdecay}\xspace}
\providecommand{\HERWIG} {{\textsc{herwig}}\xspace}
\providecommand{\HIGLU} {{\textsc{higlu}}\xspace}
\providecommand{\HIJING} {{\textsc{hijing}}\xspace}
\providecommand{\IGUANA} {\textsc{iguana}\xspace}
\providecommand{\ISAJET} {{\textsc{isajet}}\xspace}
\providecommand{\ISAPYTHIA} {{\textsc{isapythia}}\xspace}
\providecommand{\ISASUGRA} {{\textsc{isasugra}}\xspace}
\providecommand{\ISASUSY} {{\textsc{isasusy}}\xspace}
\providecommand{\ISAWIG} {{\textsc{isawig}}\xspace}
\providecommand{\MADGRAPH} {\textsc{MadGraph}\xspace}
\providecommand{\MCATNLO} {\textsc{mc@nlo}\xspace}
\providecommand{\MCFM} {\textsc{mcfm}\xspace}
\providecommand{\MILLEPEDE} {{\textsc{millepede}}\xspace}
\providecommand{\ORCA} {{\textsc{orca}}\xspace}
\providecommand{\OSCAR} {{\textsc{oscar}}\xspace}
\providecommand{\PHOTOS} {\textsc{photos}\xspace}
\providecommand{\PROSPINO} {\textsc{prospino}\xspace}
\providecommand{\PYTHIA} {{\textsc{pythia}}\xspace}
\providecommand{\SHERPA} {{\textsc{sherpa}}\xspace}
\providecommand{\TAUOLA} {\textsc{tauola}\xspace}
\providecommand{\TOPREX} {\textsc{TopReX}\xspace}
\providecommand{\XDAQ} {{\textsc{xdaq}}\xspace}

%  Experiments
\newcommand {\DZERO}{D\O\xspace}     %etc.

% Measurements and units...

\newcommand{\de}{\ensuremath{^\circ}}
\newcommand{\ten}[1]{\ensuremath{\times \text{10}^\text{#1}}}
\newcommand{\unit}[1]{\ensuremath{\text{\,#1}}\xspace}
\newcommand{\mum}{\ensuremath{\,\mu\text{m}}\xspace}
\newcommand{\micron}{\ensuremath{\,\mu\text{m}}\xspace}
\newcommand{\cm}{\ensuremath{\,\text{cm}}\xspace}
\newcommand{\mm}{\ensuremath{\,\text{mm}}\xspace}
\newcommand{\mus}{\ensuremath{\,\mu\text{s}}\xspace}
\newcommand{\keV}{\ensuremath{\,\text{ke\hspace{-.08em}V}}\xspace}
\newcommand{\MeV}{\ensuremath{\,\text{Me\hspace{-.08em}V}}\xspace}
\newcommand{\GeV}{\ensuremath{\,\text{Ge\hspace{-.08em}V}}\xspace}
\newcommand{\TeV}{\ensuremath{\,\text{Te\hspace{-.08em}V}}\xspace}
\newcommand{\PeV}{\ensuremath{\,\text{Pe\hspace{-.08em}V}}\xspace}
\newcommand{\keVc}{\ensuremath{{\,\text{ke\hspace{-.08em}V\hspace{-0.16em}/\hspace{-0.08em}c}}}\xspace}
\newcommand{\MeVc}{\ensuremath{{\,\text{Me\hspace{-.08em}V\hspace{-0.16em}/\hspace{-0.08em}c}}}\xspace}
\newcommand{\GeVc}{\ensuremath{{\,\text{Ge\hspace{-.08em}V\hspace{-0.16em}/\hspace{-0.08em}c}}}\xspace}
\newcommand{\TeVc}{\ensuremath{{\,\text{Te\hspace{-.08em}V\hspace{-0.16em}/\hspace{-0.08em}c}}}\xspace}
\newcommand{\keVcc}{\ensuremath{{\,\text{ke\hspace{-.08em}V\hspace{-0.16em}/\hspace{-0.08em}c}^\text{2}}}\xspace}
\newcommand{\MeVcc}{\ensuremath{{\,\text{Me\hspace{-.08em}V\hspace{-0.16em}/\hspace{-0.08em}c}^\text{2}}}\xspace}
\newcommand{\GeVcc}{\ensuremath{{\,\text{Ge\hspace{-.08em}V\hspace{-0.16em}/\hspace{-0.08em}c}^\text{2}}}\xspace}
\newcommand{\TeVcc}{\ensuremath{{\,\text{Te\hspace{-.08em}V\hspace{-0.16em}/\hspace{-0.08em}c}^\text{2}}}\xspace}

\newcommand{\pbinv} {\mbox{\ensuremath{\,\text{pb}^\text{$-$1}}}\xspace}
\newcommand{\fbinv} {\mbox{\ensuremath{\,\text{fb}^\text{$-$1}}}\xspace}
\newcommand{\nbinv} {\mbox{\ensuremath{\,\text{nb}^\text{$-$1}}}\xspace}
\newcommand{\percms}{\ensuremath{\,\text{cm}^\text{$-$2}\,\text{s}^\text{$-$1}}\xspace}
\newcommand{\lumi}{\ensuremath{\mathcal{L}}\xspace}
\newcommand{\Lumi}{\ensuremath{\mathcal{L}}\xspace}%both upper and lower
%
% Need a convention here:
\newcommand{\LvLow}  {\ensuremath{\mathcal{L}=\text{10}^\text{32}\,\text{cm}^\text{$-$2}\,\text{s}^\text{$-$1}}\xspace}
\newcommand{\LLow}   {\ensuremath{\mathcal{L}=\text{10}^\text{33}\,\text{cm}^\text{$-$2}\,\text{s}^\text{$-$1}}\xspace}
\newcommand{\lowlumi}{\ensuremath{\mathcal{L}=\text{2}\times \text{10}^\text{33}\,\text{cm}^\text{$-$2}\,\text{s}^\text{$-$1}}\xspace}
\newcommand{\LMed}   {\ensuremath{\mathcal{L}=\text{2}\times \text{10}^\text{33}\,\text{cm}^\text{$-$2}\,\text{s}^\text{$-$1}}\xspace}
\newcommand{\LHigh}  {\ensuremath{\mathcal{L}=\text{10}^\text{34}\,\text{cm}^\text{-2}\,\text{s}^\text{$-$1}}\xspace}
\newcommand{\hilumi} {\ensuremath{\mathcal{L}=\text{10}^\text{34}\,\text{cm}^\text{-2}\,\text{s}^\text{$-$1}}\xspace}

% Some usual physics terms

\newcommand{\zp}{\ensuremath{\mathrm{Z}^\prime}\xspace}

% SM (still to be classified)

\newcommand{\kt}{\ensuremath{k_{\mathrm{T}}}\xspace}
\newcommand{\BC}{\ensuremath{{B_{\mathrm{c}}}}\xspace}
\newcommand{\bbarc}{\ensuremath{{\overline{b}c}}\xspace}
\newcommand{\bbbar}{\ensuremath{{b\overline{b}}}\xspace}
\newcommand{\ccbar}{\ensuremath{{c\overline{c}}}\xspace}
\newcommand{\JPsi}{\ensuremath{{J}/\psi}\xspace}
\newcommand{\bspsiphi}{\ensuremath{B_s \to \JPsi\, \phi}\xspace}
\newcommand{\AFB}{\ensuremath{A_\mathrm{FB}}\xspace}
\newcommand{\EE}{\ensuremath{e^+e^-}\xspace}
\newcommand{\MM}{\ensuremath{\mu^+\mu^-}\xspace}
\newcommand{\TT}{\ensuremath{\tau^+\tau^-}\xspace}
\newcommand{\wangle}{\ensuremath{\sin^{2}\theta_{\mathrm{eff}}^\mathrm{lept}(M^2_\mathrm{Z})}\xspace}
\newcommand{\ttbar}{\ensuremath{{t\overline{t}}}\xspace}
\newcommand{\stat}{\ensuremath{\,\text{(stat.)}}\xspace}
\newcommand{\syst}{\ensuremath{\,\text{(syst.)}}\xspace}
% these moved to similar defs
%\newcommand{\Etmiss}{\ensuremath{E_{\mathrm{T}\!{\rm miss}}}}
%\newcommand{\VEtmiss}{\ensuremath{{\vec E}_{\mathrm{T}\!{\rm miss}}}}

%%%  E-gamma definitions
\newcommand{\HGG}{\ensuremath{\mathrm{H}\to\gamma\gamma}}
\newcommand{\gev}{\GeV}
\newcommand{\GAMJET}{\ensuremath{\gamma + \mathrm{jet}}}
\newcommand{\PPTOJETS}{\ensuremath{\mathrm{pp}\to\mathrm{jets}}}
\newcommand{\PPTOGG}{\ensuremath{\mathrm{pp}\to\gamma\gamma}}
\newcommand{\PPTOGAMJET}{\ensuremath{\mathrm{pp}\to\gamma +
\mathrm{jet}
}}
\newcommand{\MH}{\ensuremath{\mathrm{M_{\mathrm{H}}}}}
\newcommand{\RNINE}{\ensuremath{\mathrm{R}_\mathrm{9}}}
\newcommand{\DR}{\ensuremath{\Delta\mathrm{R}}}

% Physics symbols ...

\newcommand{\PT}{\ensuremath{p_{\mathrm{T}}}\xspace}
\newcommand{\pt}{\ensuremath{p_{\mathrm{T}}}\xspace}
\newcommand{\ET}{\ensuremath{E_{\mathrm{T}}}\xspace}
\newcommand{\HT}{\ensuremath{H_{\mathrm{T}}}\xspace}
\newcommand{\et}{\ensuremath{E_{\mathrm{T}}}\xspace}
\newcommand{\Em}{\ensuremath{E\!\!\!/}\xspace}
\newcommand{\Pm}{\ensuremath{p\!\!\!/}\xspace}
\newcommand{\PTm}{\ensuremath{{p\!\!\!/}_{\mathrm{T}}}\xspace}
\newcommand{\ETm}{\ensuremath{E_{\mathrm{T}}^{\mathrm{miss}}}\xspace}
\newcommand{\MET}{\ensuremath{E_{\mathrm{T}}^{\mathrm{miss}}}\xspace}
\newcommand{\ETmiss}{\ensuremath{E_{\mathrm{T}}^{\mathrm{miss}}}\xspace}
\newcommand{\VEtmiss}{\ensuremath{{\vec E}_{\mathrm{T}}^{\mathrm{miss}}}\xspace}

%%%%%%
% From Albert
%

\newcommand{\ga}{\ensuremath{\gtrsim}}
\newcommand{\la}{\ensuremath{\lesssim}}
\newcommand{\swsq}{\ensuremath{\sin^2\theta_W}\xspace}
\newcommand{\cwsq}{\ensuremath{\cos^2\theta_W}\xspace}
\newcommand{\tanb}{\ensuremath{\tan\beta}\xspace}
\newcommand{\tanbsq}{\ensuremath{\tan^{2}\beta}\xspace}
\newcommand{\sidb}{\ensuremath{\sin 2\beta}\xspace}
\newcommand{\alpS}{\ensuremath{\alpha_S}\xspace}
\newcommand{\alpt}{\ensuremath{\tilde{\alpha}}\xspace}

\newcommand{\QL}{\ensuremath{Q_L}\xspace}
\newcommand{\sQ}{\ensuremath{\tilde{Q}}\xspace}
\newcommand{\sQL}{\ensuremath{\tilde{Q}_L}\xspace}
\newcommand{\ULC}{\ensuremath{U_L^C}\xspace}
\newcommand{\sUC}{\ensuremath{\tilde{U}^C}\xspace}
\newcommand{\sULC}{\ensuremath{\tilde{U}_L^C}\xspace}
\newcommand{\DLC}{\ensuremath{D_L^C}\xspace}
\newcommand{\sDC}{\ensuremath{\tilde{D}^C}\xspace}
\newcommand{\sDLC}{\ensuremath{\tilde{D}_L^C}\xspace}
\newcommand{\LL}{\ensuremath{L_L}\xspace}
\newcommand{\sL}{\ensuremath{\tilde{L}}\xspace}
\newcommand{\sLL}{\ensuremath{\tilde{L}_L}\xspace}
\newcommand{\ELC}{\ensuremath{E_L^C}\xspace}
\newcommand{\sEC}{\ensuremath{\tilde{E}^C}\xspace}
\newcommand{\sELC}{\ensuremath{\tilde{E}_L^C}\xspace}
\newcommand{\sEL}{\ensuremath{\tilde{E}_L}\xspace}
\newcommand{\sER}{\ensuremath{\tilde{E}_R}\xspace}
\newcommand{\sFer}{\ensuremath{\tilde{f}}\xspace}
\newcommand{\sQua}{\ensuremath{\tilde{q}}\xspace}
\newcommand{\sUp}{\ensuremath{\tilde{u}}\xspace}
\newcommand{\suL}{\ensuremath{\tilde{u}_L}\xspace}
\newcommand{\suR}{\ensuremath{\tilde{u}_R}\xspace}
\newcommand{\sDw}{\ensuremath{\tilde{d}}\xspace}
\newcommand{\sdL}{\ensuremath{\tilde{d}_L}\xspace}
\newcommand{\sdR}{\ensuremath{\tilde{d}_R}\xspace}
\newcommand{\sTop}{\ensuremath{\tilde{t}}\xspace}
\newcommand{\stL}{\ensuremath{\tilde{t}_L}\xspace}
\newcommand{\stR}{\ensuremath{\tilde{t}_R}\xspace}
\newcommand{\stone}{\ensuremath{\tilde{t}_1}\xspace}
\newcommand{\sttwo}{\ensuremath{\tilde{t}_2}\xspace}
\newcommand{\sBot}{\ensuremath{\tilde{b}}\xspace}
\newcommand{\sbL}{\ensuremath{\tilde{b}_L}\xspace}
\newcommand{\sbR}{\ensuremath{\tilde{b}_R}\xspace}
\newcommand{\sbone}{\ensuremath{\tilde{b}_1}\xspace}
\newcommand{\sbtwo}{\ensuremath{\tilde{b}_2}\xspace}
\newcommand{\sLep}{\ensuremath{\tilde{l}}\xspace}
\newcommand{\sLepC}{\ensuremath{\tilde{l}^C}\xspace}
\newcommand{\sEl}{\ensuremath{\tilde{e}}\xspace}
\newcommand{\sElC}{\ensuremath{\tilde{e}^C}\xspace}
\newcommand{\seL}{\ensuremath{\tilde{e}_L}\xspace}
\newcommand{\seR}{\ensuremath{\tilde{e}_R}\xspace}
\newcommand{\snL}{\ensuremath{\tilde{\nu}_L}\xspace}
\newcommand{\sMu}{\ensuremath{\tilde{\mu}}\xspace}
\newcommand{\sNu}{\ensuremath{\tilde{\nu}}\xspace}
\newcommand{\sTau}{\ensuremath{\tilde{\tau}}\xspace}
\newcommand{\Glu}{\ensuremath{g}\xspace}
\newcommand{\sGlu}{\ensuremath{\tilde{g}}\xspace}
\newcommand{\Wpm}{\ensuremath{W^{\pm}}\xspace}
\newcommand{\sWpm}{\ensuremath{\tilde{W}^{\pm}}\xspace}
\newcommand{\Wz}{\ensuremath{W^{0}}\xspace}
\newcommand{\sWz}{\ensuremath{\tilde{W}^{0}}\xspace}
\newcommand{\sWino}{\ensuremath{\tilde{W}}\xspace}
\newcommand{\Bz}{\ensuremath{B^{0}}\xspace}
\newcommand{\sBz}{\ensuremath{\tilde{B}^{0}}\xspace}
\newcommand{\sBino}{\ensuremath{\tilde{B}}\xspace}
\newcommand{\Zz}{\ensuremath{Z^{0}}\xspace}
\newcommand{\sZino}{\ensuremath{\tilde{Z}^{0}}\xspace}
\newcommand{\sGam}{\ensuremath{\tilde{\gamma}}\xspace}
\newcommand{\chiz}{\ensuremath{\tilde{\chi}^{0}}\xspace}
\newcommand{\chip}{\ensuremath{\tilde{\chi}^{+}}\xspace}
\newcommand{\chim}{\ensuremath{\tilde{\chi}^{-}}\xspace}
\newcommand{\chipm}{\ensuremath{\tilde{\chi}^{\pm}}\xspace}
\newcommand{\Hone}{\ensuremath{H_{d}}\xspace}
\newcommand{\sHone}{\ensuremath{\tilde{H}_{d}}\xspace}
\newcommand{\Htwo}{\ensuremath{H_{u}}\xspace}
\newcommand{\sHtwo}{\ensuremath{\tilde{H}_{u}}\xspace}
\newcommand{\sHig}{\ensuremath{\tilde{H}}\xspace}
\newcommand{\sHa}{\ensuremath{\tilde{H}_{a}}\xspace}
\newcommand{\sHb}{\ensuremath{\tilde{H}_{b}}\xspace}
\newcommand{\sHpm}{\ensuremath{\tilde{H}^{\pm}}\xspace}
\newcommand{\hz}{\ensuremath{h^{0}}\xspace}
\newcommand{\Hz}{\ensuremath{H^{0}}\xspace}
\newcommand{\Az}{\ensuremath{A^{0}}\xspace}
\newcommand{\Hpm}{\ensuremath{H^{\pm}}\xspace}
\newcommand{\sGra}{\ensuremath{\tilde{G}}\xspace}
\newcommand{\mtil}{\ensuremath{\tilde{m}}\xspace}
\newcommand{\rpv}{\ensuremath{\rlap{\kern.2em/}R}\xspace}
\newcommand{\LLE}{\ensuremath{LL\bar{E}}\xspace}
\newcommand{\LQD}{\ensuremath{LQ\bar{D}}\xspace}
\newcommand{\UDD}{\ensuremath{\overline{UDD}}\xspace}
\newcommand{\Lam}{\ensuremath{\lambda}\xspace}
\newcommand{\Lamp}{\ensuremath{\lambda'}\xspace}
\newcommand{\Lampp}{\ensuremath{\lambda''}\xspace}
\newcommand{\spinbd}[2]{\ensuremath{\bar{#1}_{\dot{#2}}}\xspace}

\newcommand{\MD}{\ensuremath{{M_\mathrm{D}}}\xspace}% ED mass
\newcommand{\Mpl}{\ensuremath{{M_\mathrm{Pl}}}\xspace}% Planck mass
\newcommand{\Rinv} {\ensuremath{{R}^{-1}}\xspace}

%%%%%%%%%%%%%%%%%%%%%%%%%%%%%%%%%%%%%%%%%%%%%%%%%%%%%%%%%%%%%%%%%%%%
%
% Hyphenations (only need to add here if you get a nasty word break)
%
\hyphenation{en-viron-men-tal}%    just an example

% --- force emacs to latex mode -*-latex-*-
%
% Tell TeX where to look for graphics files to be included
%
\graphicspath{{pdf/}} % Note, the additional {} are mandatory!

%
% Put additional commands, abbreviations etc. here 
%
\providecommand{\FIXME}[1]{({\bf FIXME: #1})}
\providecommand{\HERWIGPP} {{\textsc{herwig++}}\xspace}
\providecommand{\NLOJETPP} {{\textsc{nlojet++}}\xspace}
\providecommand{\Et}{E_{\mathrm{T}}}
\providecommand{\met}{\mbox{${\hbox{$\vec{E}$\kern-0.5em\lower-.1ex\hbox{/}}}_T~$}}
\providecommand{\MET}{\mbox{${\hbox{$E$\kern-0.5em\lower-.1ex\hbox{/}}}_T~$}}
\providecommand{\pthat}{\ensuremath{\hat{\text{p}}_\mathrm{T}}\xspace}
\providecommand{\kthat}{\ensuremath{\hat{\text{k}}_\mathrm{T}}\xspace}
\providecommand{\xt}{\ensuremath{\text{x}_\mathrm{T}}\xspace}
\providecommand{\ptjet}{\ensuremath{p_{\mathrm{T,jet}}}\xspace}
\providecommand{\FASTNLO} {{fast\textsc{nlo}}\xspace}

\providecommand{\rbthm}{\rule[-2ex]{0ex}{5ex}}
\providecommand{\rbthr}{\rule[-1.7ex]{0ex}{5ex}}
\providecommand{\rbtrm}{\rule[-2ex]{0ex}{5ex}}
\providecommand{\rbtrr}{\rule[-0.8ex]{0ex}{3.2ex}}
\providecommand{\relmet}{\ensuremath{\text{MET}/\sum{E_{\mathrm{T}}}}\xspace}
\providecommand{\LvStartup}{\ensuremath{\mathcal{L}=\text{10}^\text{31}\,\text{cm}^\text{$-$2}\,\text{s}^\text{$-$1}}\xspace}

% ==============================================================================
% title page for few authors

\begin{titlepage}

  % select one of the following and type in the proper number:
  \conferencereport{2009/189} \date{15 July 2009 (v3, 19 July 2009)}
  
  \title{QCD Physics Potential of CMS}

  \begin{Authlist}
    K.~Rabbertz\\on behalf of the CMS Collaboration,
    \Instfoot{IEKP}{Institut f\"ur Experimentelle Kernphysik,
      University of Karlsruhe, Germany}
  \end{Authlist}
  
  % if needed, use the following: \collaboration{CMS collaboration}

  \begin{abstract}
    In view of the approaching LHC operation the feasibility and
    accuracy of QCD measurements with the CMS experiment at the Large
    Hadron Collider (LHC) involving hadrons and jets are discussed.
    This summary is based on analyses performed at CMS for
    center-of-mass energies of $10$ as well as $14\,{\rm TeV}$
    assuming event numbers ranging from some days of data taking up to
    $100\,{\rm pb}^{-1}$ of integrated luminosity with proton-proton
    collisions.
  \end{abstract}

  % if needed, use the following:
  \conference{Presented at {\it 1st IPM Meeting on LHC Physics},
    Isfahan, Iran, 20.-24. April 2009}
%  \note{Almost final version}
  
\end{titlepage}

\setcounter{page}{2}%JPP

\section{Introduction}

With the advent of the LHC, a completely new regime in centre-of-mass
energy for hadron-hadron collisions will be explored. While the main
interest of the LHC is to unravel the nature of electroweak symmetry
breaking, a detailed understanding of the detector performance and the
Standard Model processes is a must.  QCD, the theory of the strong
interaction, describes one of the four fundamental forces of nature
and in particular the hard interactions between coloured
quarks and gluons and how they bind together to form hadrons.  Due to
the huge cross sections of QCD reactions involved, the most
outstanding feature of events at the \TeV energy scale is therefore
the abundant production of jets, i.e.\ collimated streams of hadrons
that are supposed to originate from a common initiator.  A profound
understanding of hadron production and jet physics therefore poses the
foundation for the physics commissioning and monitoring of the CMS
experiment~\cite{Adolphi:2008zzk} and is a mandatory step in order to
re-establish the Standard Model and to set the stage for the search of
new phenomena.

In the next section analyses dealing with first measurements of hadron
production and of the Underlying Event (UE) activity based on jets formed
from tracks of charged particles are presented.  The following section
then concentrates on jet physics
% , jet substructure and event shapes
before finishing this report with an outlook.

Photon physics which in CMS is included in the QCD working group as
well had to be left out. The latest results from CMS can be found
in~\cite{Ball:2007zza} and~\cite{Gupta:2008zza}.  Details on the
performance of the CMS experiment with respect to track
reconstruction, alignment, jet finding and calibration can be found
elsewhere in these proceedings.

\section{Tracks and Hadrons}
\subsection{Charged Hadron Production}

Charged particle multiplicity distributions from hadron-hadron
collisions have been studied already in other
experiments~\cite{Alner:1986xu,Abe:1989td}.  A measurement of the
distribution in pseudorapidity $\eta=-\ln\tan(\theta/2)$ with $\theta$
being the polar angle, $dN_{ch}/d\eta$, can be carried out with a few
thousand events collected by the CMS detector and will be one of the
first measurements at the LHC. Since one has to integrate over the
transverse momentum spectrum for each pseudorapidity region, however,
one needs to extrapolate the measurable \pt range to small momenta due
to the limitation of track finding in the low \pt limit. To reduce
unavoidable modeling systematics three different methods to reach as
low in \pt as possible are foreseen in
CMS~\cite{CMS-PAS-QCD-08-004,CMS-PAS-QCD-07-001,CMS-PAS-QCD-09-002}.

\begin{figure}[htbp]
  \centering
  \includegraphics[width=0.30\textwidth]{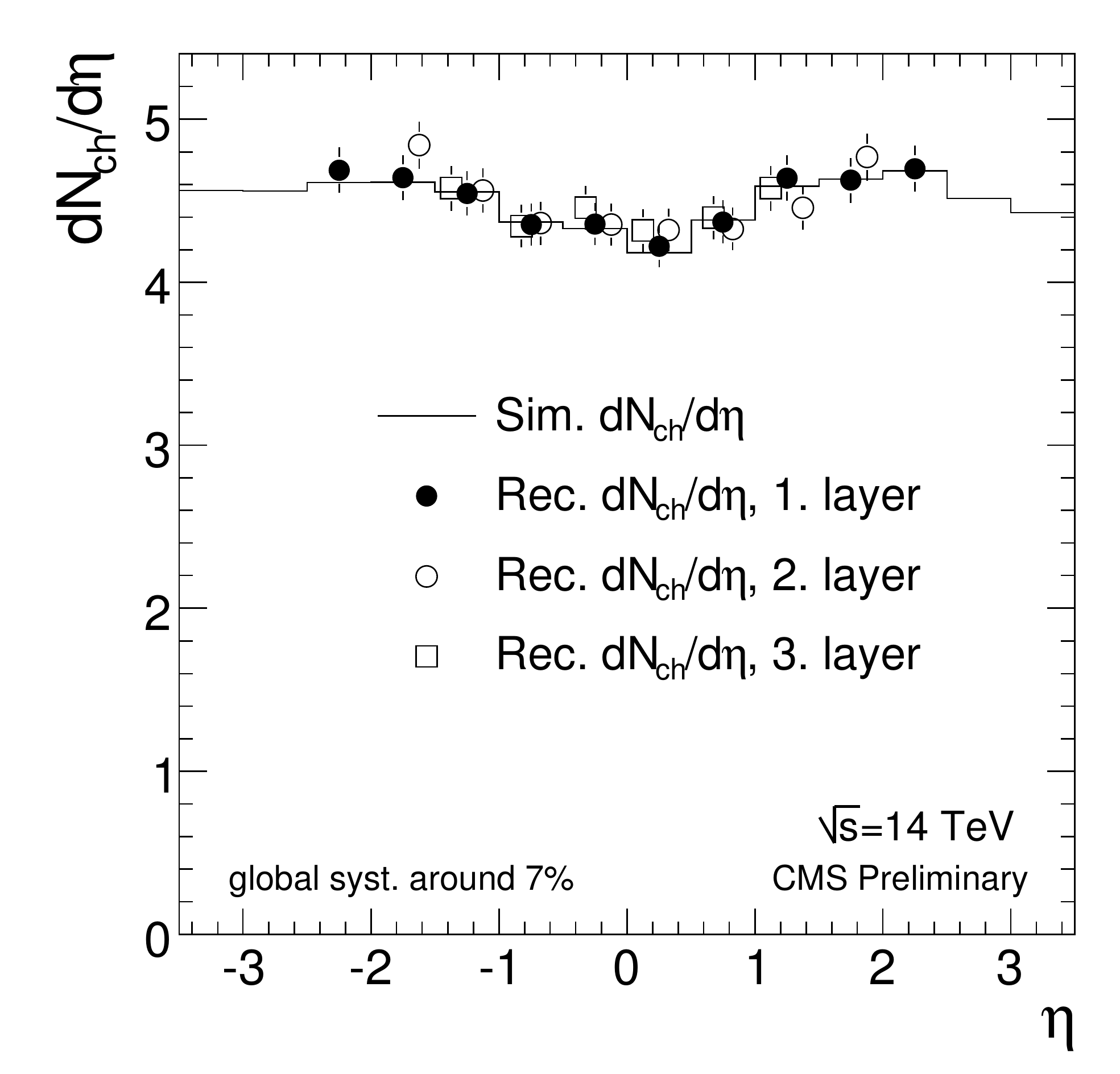}%
  \includegraphics[width=0.33\textwidth]{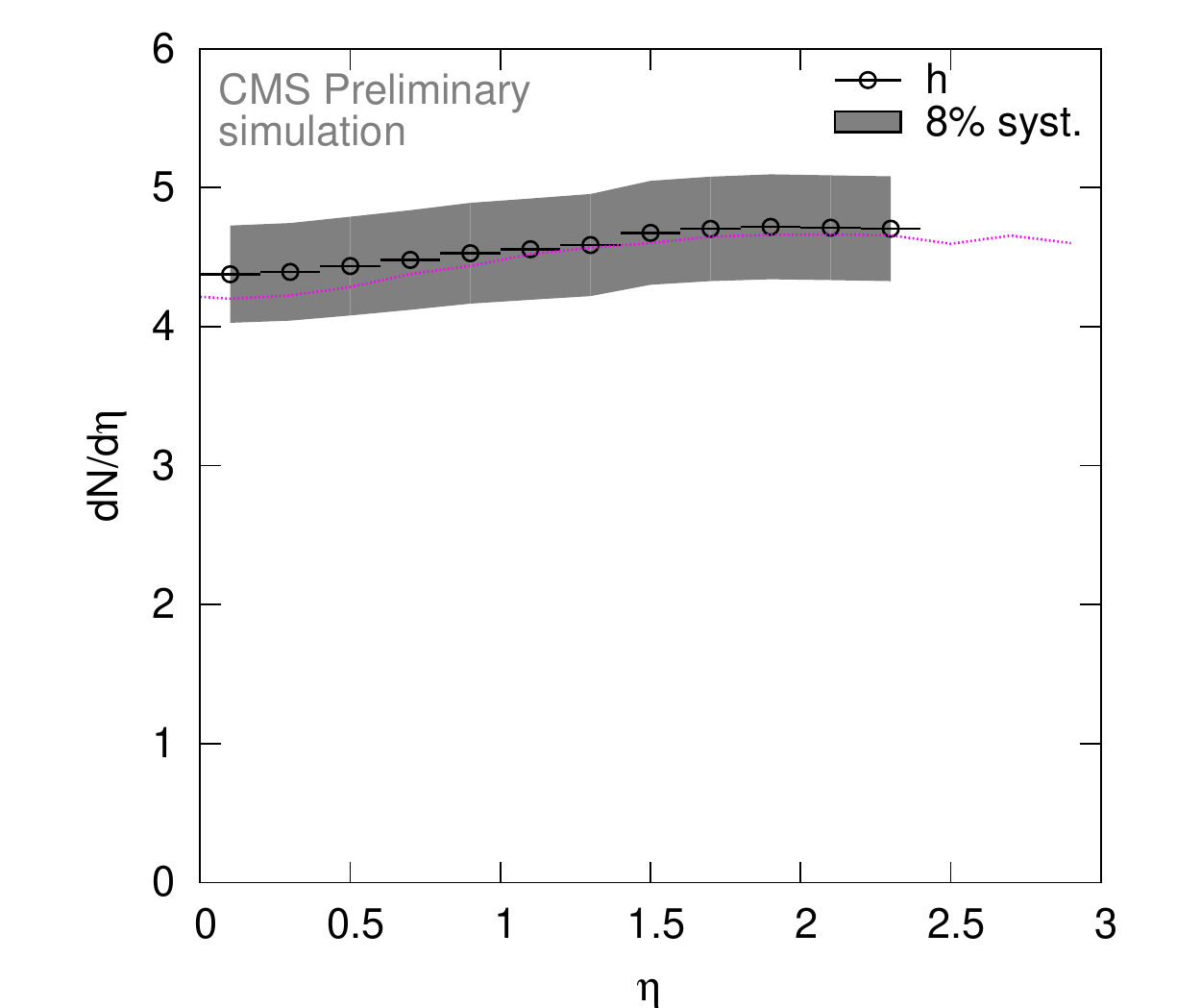}%
  \includegraphics[width=0.31\textwidth]{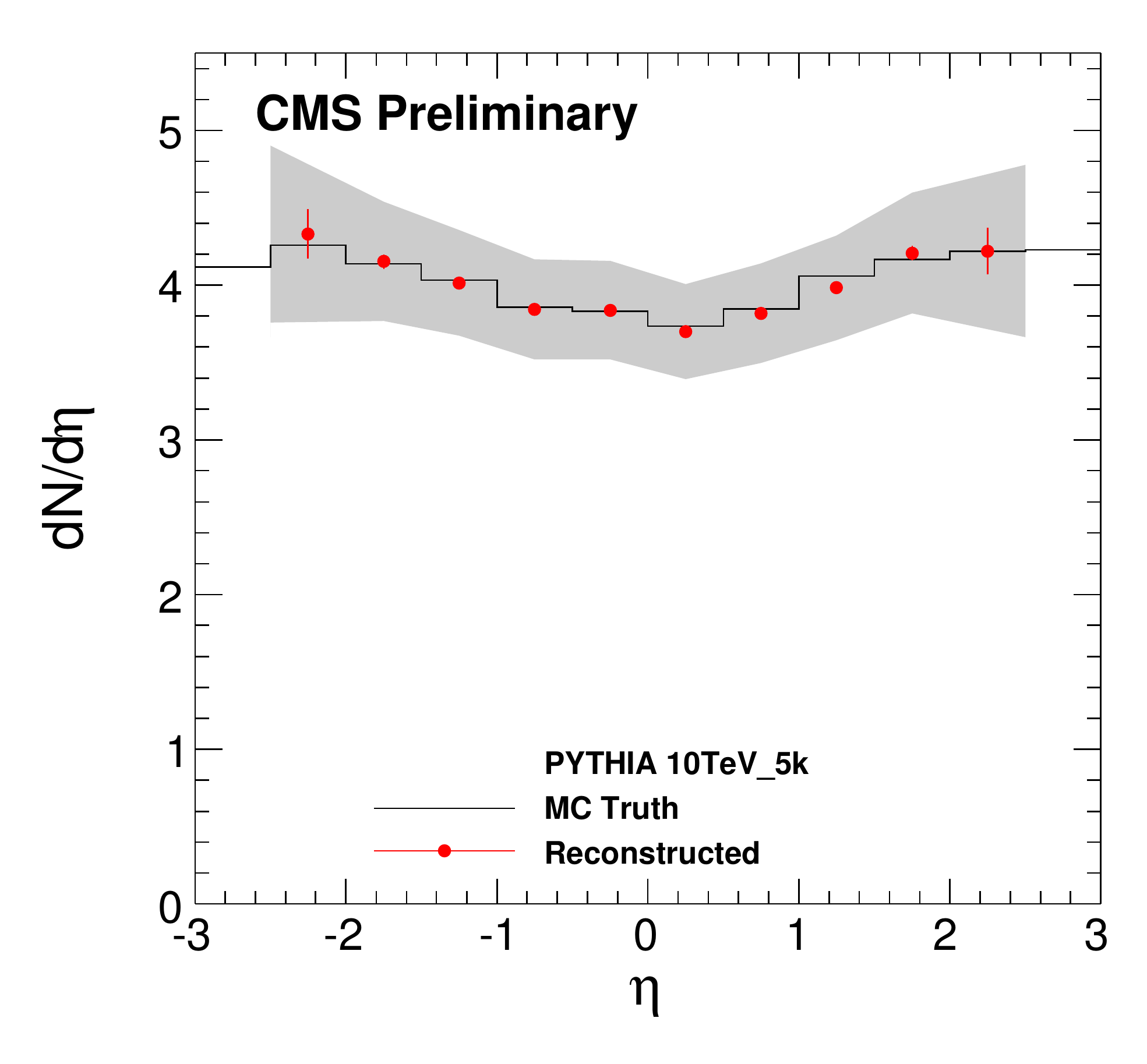}
  \caption{Simulation results for charged particle densities in
    pseudorapidity using the methods of pixel hit counting (left,
    $\sqrt{s}=14\TeV$), pixel tracklets (middle, $\sqrt{s}=10\TeV$)
    and track reconstruction with pixel triplets (right,
    $\sqrt{s}=10\TeV$) including estimates of the systematic
    uncertainties of $\approx 7-10\%$ are shown together with input
    predictions from \PYTHIA.}
  \label{fig:dNdeta}
\end{figure}

The first consists of a hit-counting technique~\cite{Back:2001bq}
where charged particles are only required to reach the first layer of
the CMS pixel detector, hence $\pt \gtrsim 30\MeV$.  The advantage of
this method is its relative insensitivity to detector misalignment,
however, it depends on details of the simulation of the pixel
response.  The results from all three pixel layers (with different
reaches in low \pt) can be compared. To reduce the sensitivity to the
detailed detector response simulation, tracklets consisting of two-hit
pixel tracks in consecutive layers are employed as suggested
in~\cite{Back:2000gw}. Finally, a track-reconstruction method with
pixel hit triplets working down to $\pt \approx 100\MeV$ is proposed
which requires a more careful study of the tracker alignment. The
three techniques exhibit different sensitivities to the diverse
sources of systematic uncertainty and complement each other.
Simulation results for all three are shown in figure~\ref{fig:dNdeta}.

\subsection{Underlying Event Measurements}

Another analysis~\cite{CMS-PAS-QCD-07-003} exploits the standard track
reconstruction for $\pt > 900\MeV$ with the silicon strip tracker of
CMS\@. After triggering on Minimum Bias or jet events with different
jet \pt thresholds, all the tracks are investigated with respect to
the difference in azimuth towards the leading jet constructed from
these tracks. In other
experiments~\cite{Affolder:2001xt,Acosta:2004wqa} it could be shown
that the transverse region of $60^\circ<|\Delta\phi|<120^\circ$ with
respect to the leading jet is most sensitive to the Underlying Event,
i.e.\ every collision product not coming directly from the hard
scatter. Extrapolations of the UE contributions to events at LHC
energies vary widely such that an early determination of its size and
the tuning of the MC generators is an important start-up measurement.

\begin{figure}[htbp]
  \centering
  \includegraphics[width=0.45\textwidth]{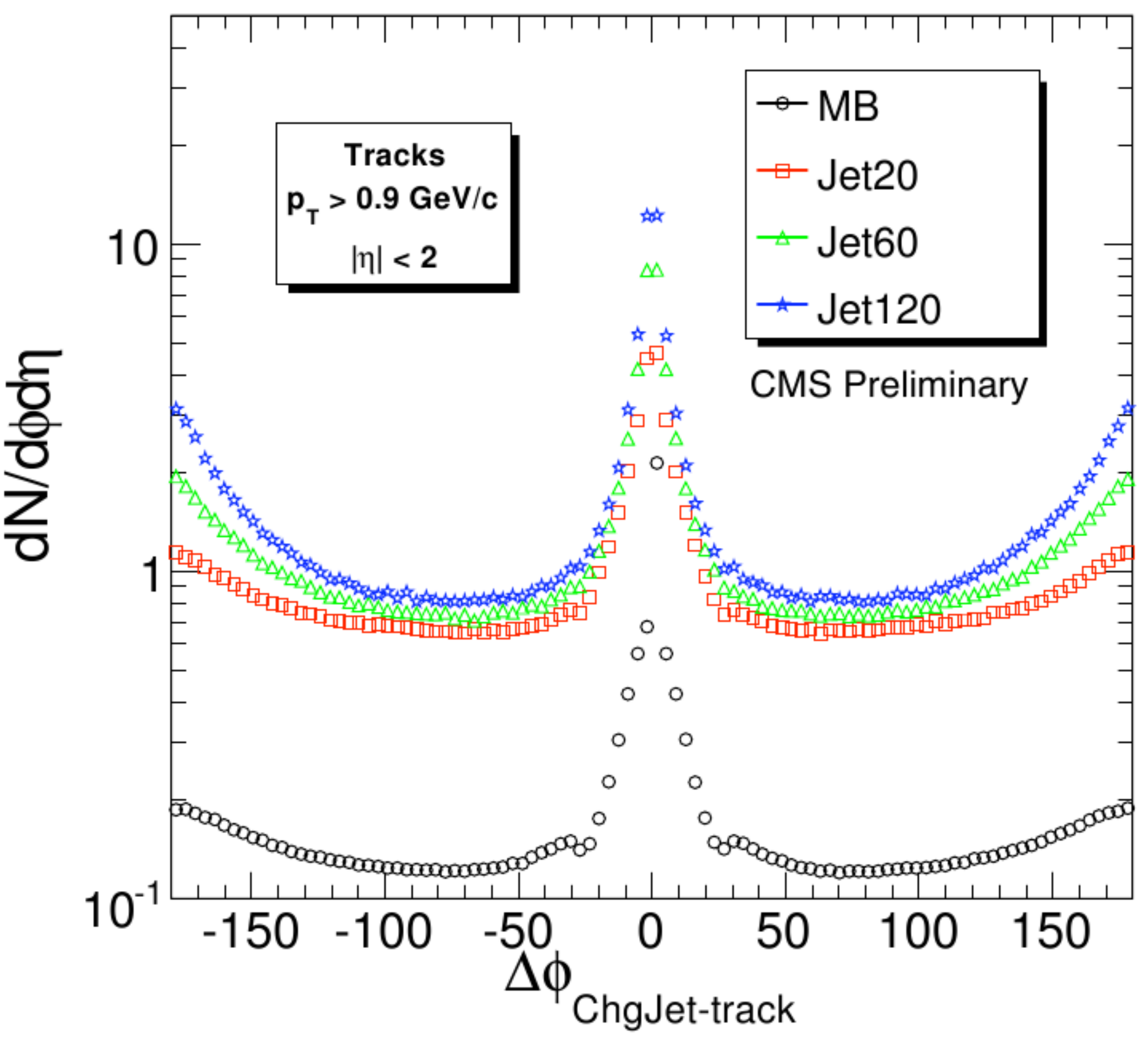}\hspace{0.08\textwidth}%
  \includegraphics[width=0.45\textwidth]{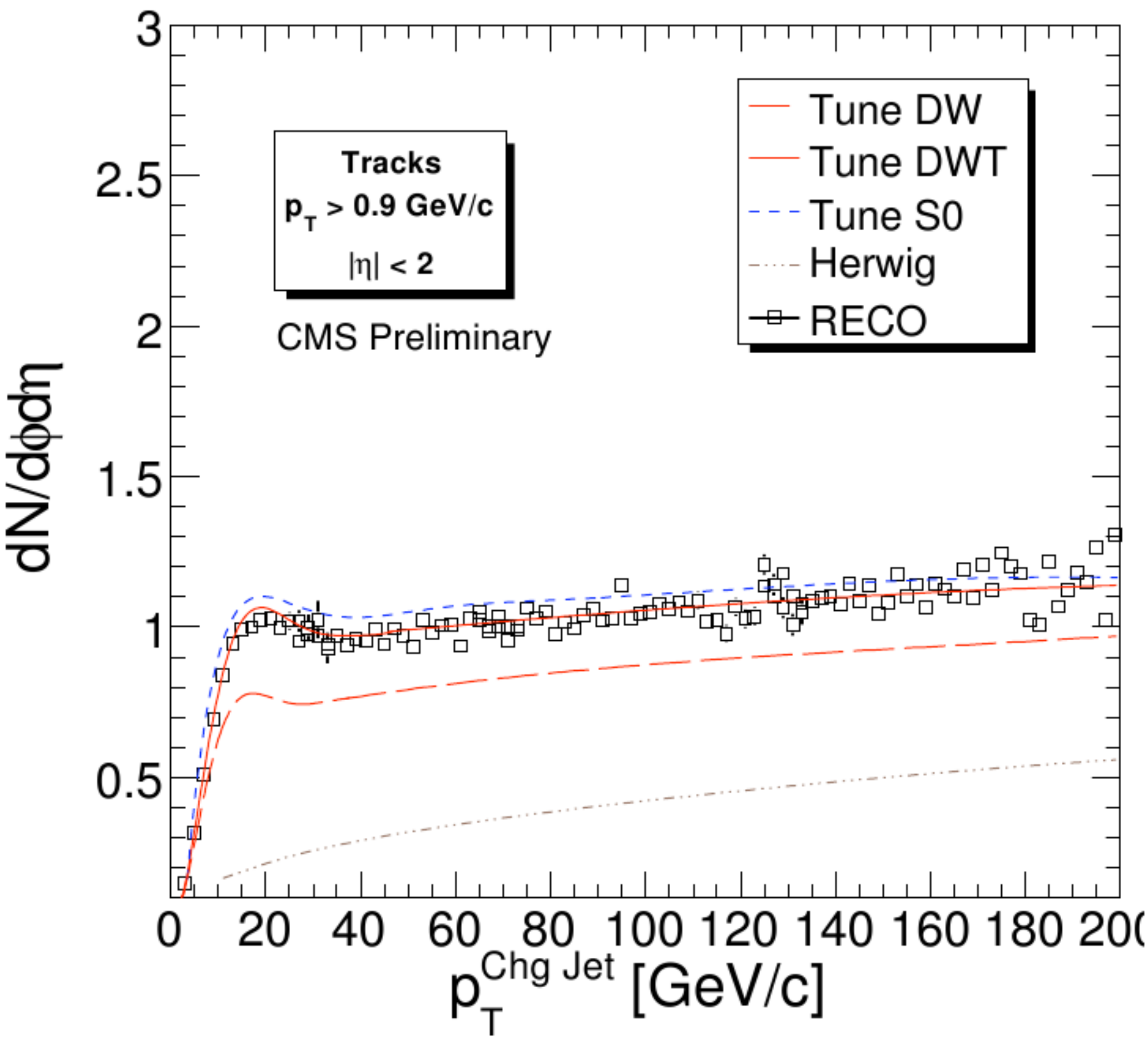}
  \caption{Composition of the total charged particle distribution in
    $\Delta\phi$ for all trigger streams (left) and the resulting \pt
    dependence of the reconstructed charged particle density in the
    transverse plane together with predictions of various \PYTHIA
    tunes and from \HERWIG assuming $10\pbinv$ of integrated
    luminosity at $\sqrt{s}=14\TeV$.}
  \label{fig:UE}
\end{figure}

Figure~\ref{fig:UE} presents the composition of the total charged
particle distribution in $\Delta\phi$ for all trigger streams on the
left and the resulting \pt dependence of the charged particle density
in the transverse plane reconstructed from simulations with \PYTHIA
tune DWT on the right. For comparison the MC predictions of \PYTHIA
with various tunes and from \HERWIG without model for multiple parton
interactions are shown as well. Already with the assumed $10\pbinv$ of
integrated luminosity at $\sqrt{s}=14\TeV$ it will be possible to
differentiate between the extrapolations of some models to LHC
energies. Using tracks with a lower limit of $\pt > 500\MeV$ the
sensitivity can be further increased as demonstrated
in~\cite{CMS-PAS-QCD-07-003}.

\vspace*{-0.15cm}
\section{Jet Physics}

In contrast to the last section where jets were used at most in order
to fix the basic orientation of an event, they are the primary topic
now. Instead of looking after global event properties like the numbers
of produced hadrons or the general flow of momentum in an event, one
would like to establish a closer connection to the hard process which
is described theoretically in terms of partons, i.e.\ quarks,
anti-quarks and gluons.  Since it is impossible to unambiguously
assign bunches of observed hadrons to the originating partons, jet
algorithms are employed that define a distance measure between objects
and uniquely determine which of them are sufficiently close to each
other to be considered to come from the same origin and hence to
combine them into a jet.

In CMS three jet algorithms with in total five different jet sizes $R$
(or $D$) are in use: The Iterative Cone algorithm with $R=0.5$ as
implemented in the trigger of the CMS experiment~\cite{Acosta:922757},
the SISCone algorithm~\cite{Salam:2007xv} with $R=0.5$ or $R=0.7$, and
the \kt algorithm~\cite{Ellis:1993tq, Catani:1992zp, Catani:1993hr}
with $D=0.4$ or $D=0.6$. For SISCone and \kt the implementation
of~\cite{Cacciari:2005hq} has been employed.  It has to be noted that
the Iterative Cone is not collinear and infrared-safe.% and therefore
% has to be avoided for serious comparisons with theory calculations
% in perturbative QCD (pQCD).

For safe jet algorithms the following theoretical uncertainties in
approximate order of importance have to be considered when comparing
pQCD results to experimental data: The uncertainty inherent in the
determination of the parton density functions (PDFs) of the proton,
the precision in perturbative QCD (leading order LO, next-to-leading
order NLO, \ldots)~\footnote{The uncertainty of a pQCD calculation is
  conventionally estimated by varying the renormalization and
  factorization scales.}, non-perturbative corrections, the dependence
on the PDF parameterizations, the knowledge of $\alpS(M_Z)$, and for
very high jet transverse momenta potentially electroweak corrections.

On the experimental side the dominant uncertainties are due to the jet
energy calibration JEC (including the treatment of electronic noise
and of collisions from different proton bunch crossings, i.e.\
pile-up), the luminosity determination, the jet energy resolution JER,
trigger efficiencies, the spatial resolutions in azimuthal angle
$\phi$ and in pseudorapidity $\eta$, and non-collision background.
% lists non-exhaustive
Depending on a particular jet analysis the sensitivity to one or
another effect might be reduced. For example in the case of normalized
observables like the dijet azimuthal decorrelation and event shapes or
in cross-section ratios like the dijet production ratio in
pseudorapidity and 3-jet to all-jet ratios, the luminosity uncertainty
is eliminated and the uncertainty due to the JEC is reduced.  The
inclusive jet cross section, which will be discussed first, is a
particularly challenging measurement and requires all uncertainties to
be under control.

\subsection{Inclusive Jets}

In~\cite{CMS-PAS-QCD-08-001} a plan for the measurement of the
differential inclusive jet production cross section for rapidities up
to $|y|=2.5$ with CMS assuming $10\pbinv$ of integrated luminosity at
a center of mass energy of $\sqrt{s}=10\TeV$ is
presented.\footnote{Forward jets with $3<|\eta|<5$ have been
  investigated in~\cite{CMS-PAS-FWD-08-001} for $\sqrt{s}=14\TeV$ .}
The reach in jet transverse momentum is already beyond any previous
collider experiment~\cite{Aaltonen:2008eq, Abulencia:2007ez,
  Abazov:2008hu} and the \TeV scale of jet physics can be probed.  The
analysis is performed on fully simulated CMS events which are adopted
as pseudo data. Jets are reconstructed from calorimeter energy
depositions with the inclusive \kt ($D=0.6$) and the SISCone ($R=0.7$)
algorithm.  Events accepted by the trigger simulation are combined to
the inclusive jet \pt spectrum in such a way that each \pt bin
receives contributions from exactly one fully efficient trigger.

\begin{figure}[htbp]
  \centering
  \includegraphics[width=0.50\textwidth]{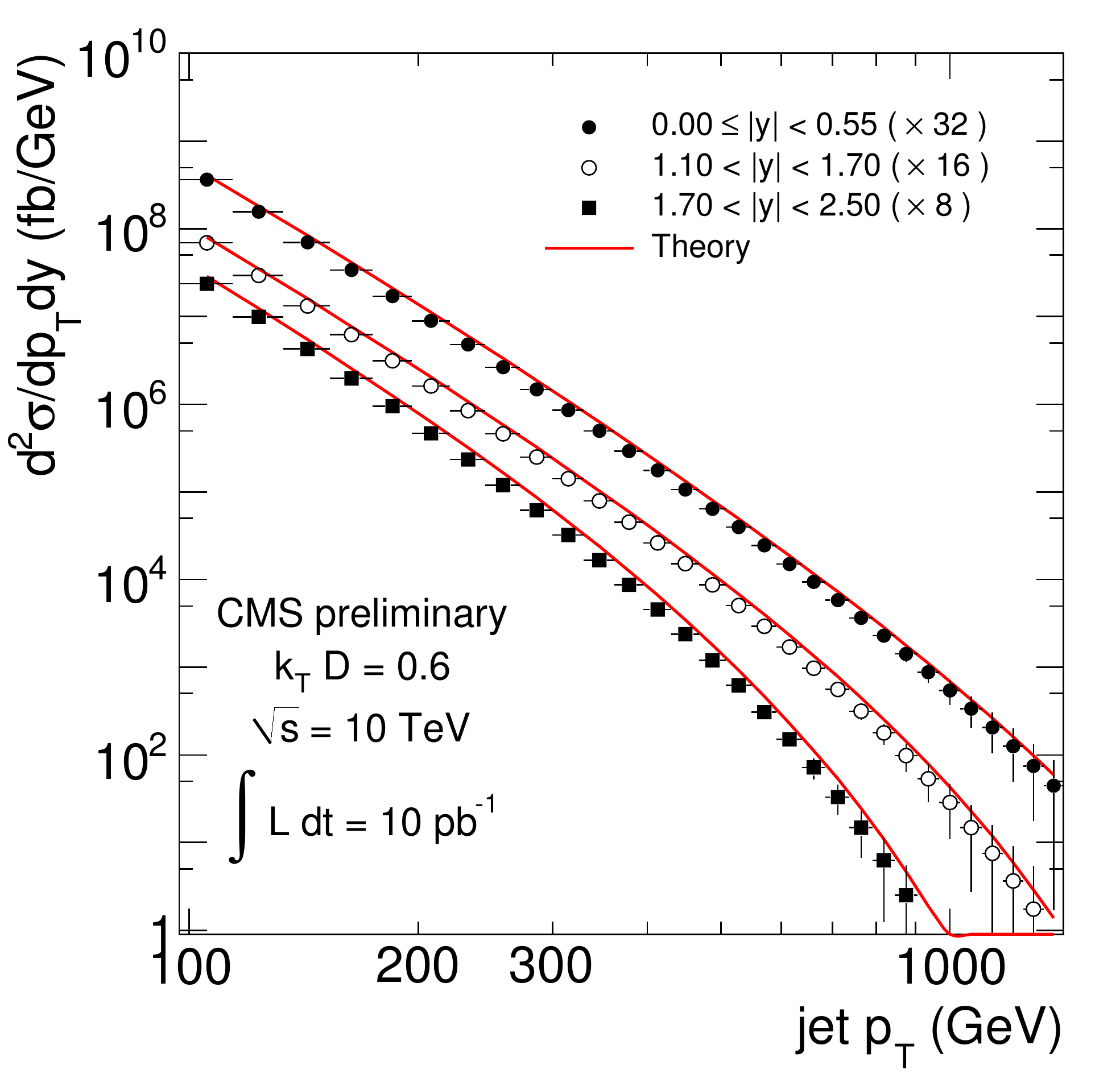}%
  \includegraphics[width=0.50\textwidth]{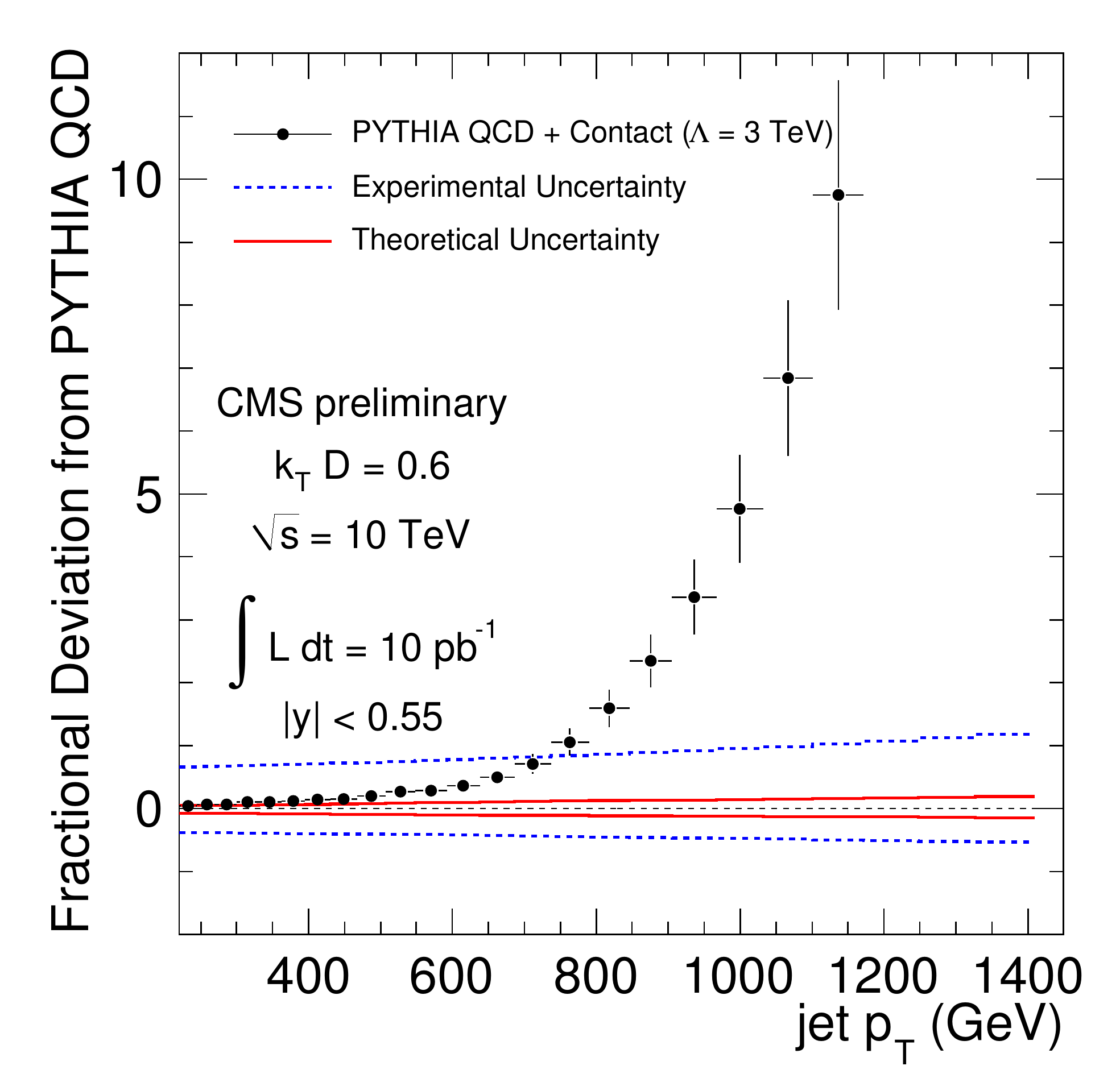}
  \caption{Comparison between the corrected measured spectra and the
    theory predictions for the \kt algorithm (left).  For better
    visibility the spectra have been multiplied by factors of 8, 16
    and 32.  Fractional difference of a \PYTHIA QCD+$3\TeV$ contact
    interaction term and pure \PYTHIA QCD in comparison to the
    experimental and theoretical uncertainties (right).}
  \label{fig:InclJetsComparison}
\end{figure}

Subsequently, each jet is subjected to a JEC that corrects on average
the observed jet energy to the energy of the final state particle
jet~\cite{CMS-PAS-JME-07-002}. Lacking collision data the JEC is
currently derived from Monte Carlo truth by matching reconstructed
jets with generated particle jets. Due to the fact that the QCD jet
\pt spectrum is steeply falling an additional unsmearing step becomes
necessary. There are more jets migrating to higher transverse momenta
than in the opposite direction because of the finite jet energy
resolution. To remove this distortion from the measured spectrum the
\textit{Ansatz Method} is used, which has been employed successfully
at the Tevatron~\cite{Abbott:2000kp,Abazov:2008hu}. The corrected
\ptjet spectra (times $K$ factors) for three regions in absolute
rapidity
$|y|=\left|\frac{1}{2}\ln\left(\frac{E+p_z}{E-p_z}\right)\right|$ are
compared to theory predictions (times non-perturbative corrections) in
figure~\ref{fig:InclJetsComparison} left.

\begin{figure}[htbp]
  \centering
  \includegraphics[width=0.50\textwidth]{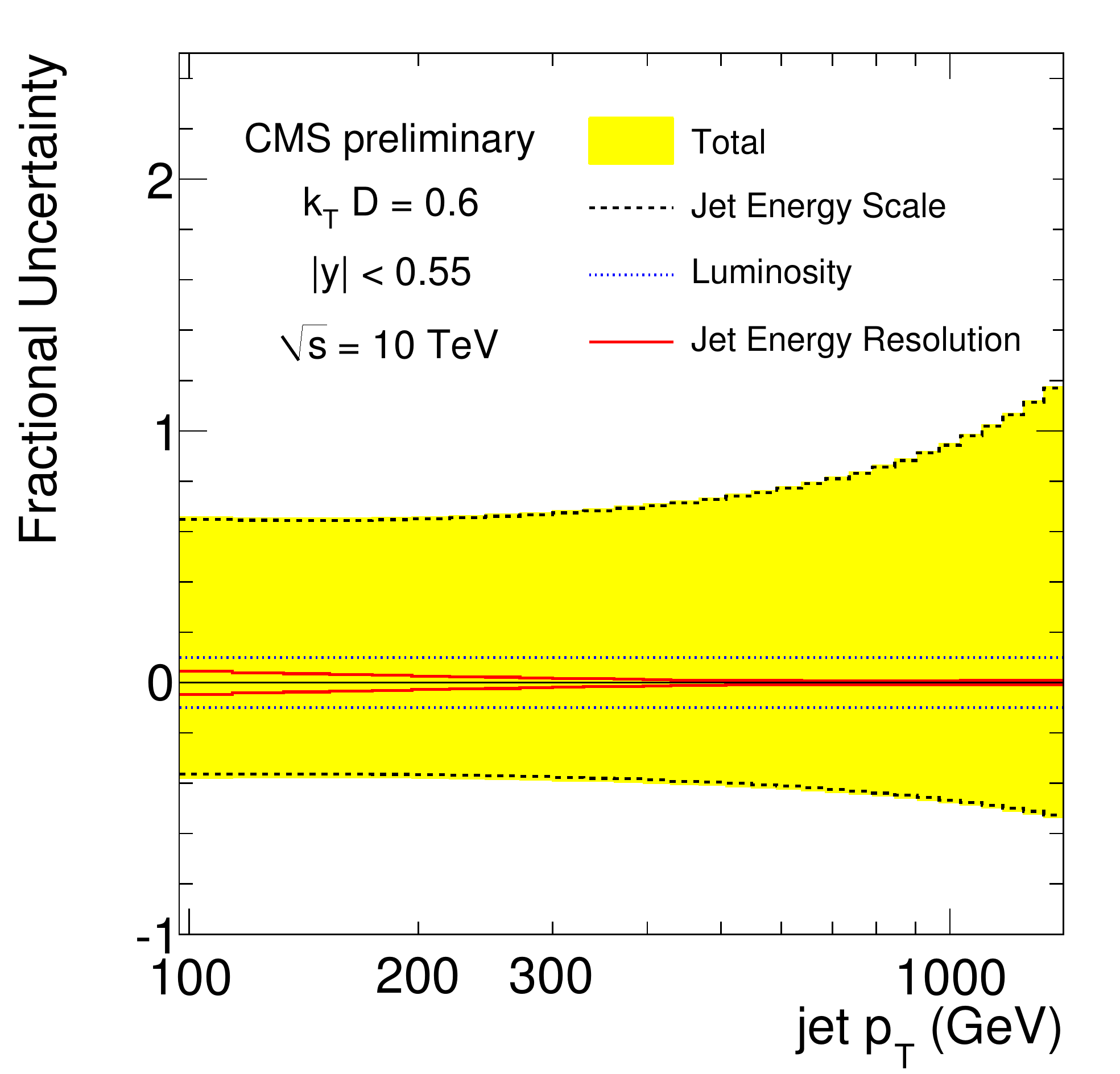}%
  \includegraphics[width=0.50\textwidth]{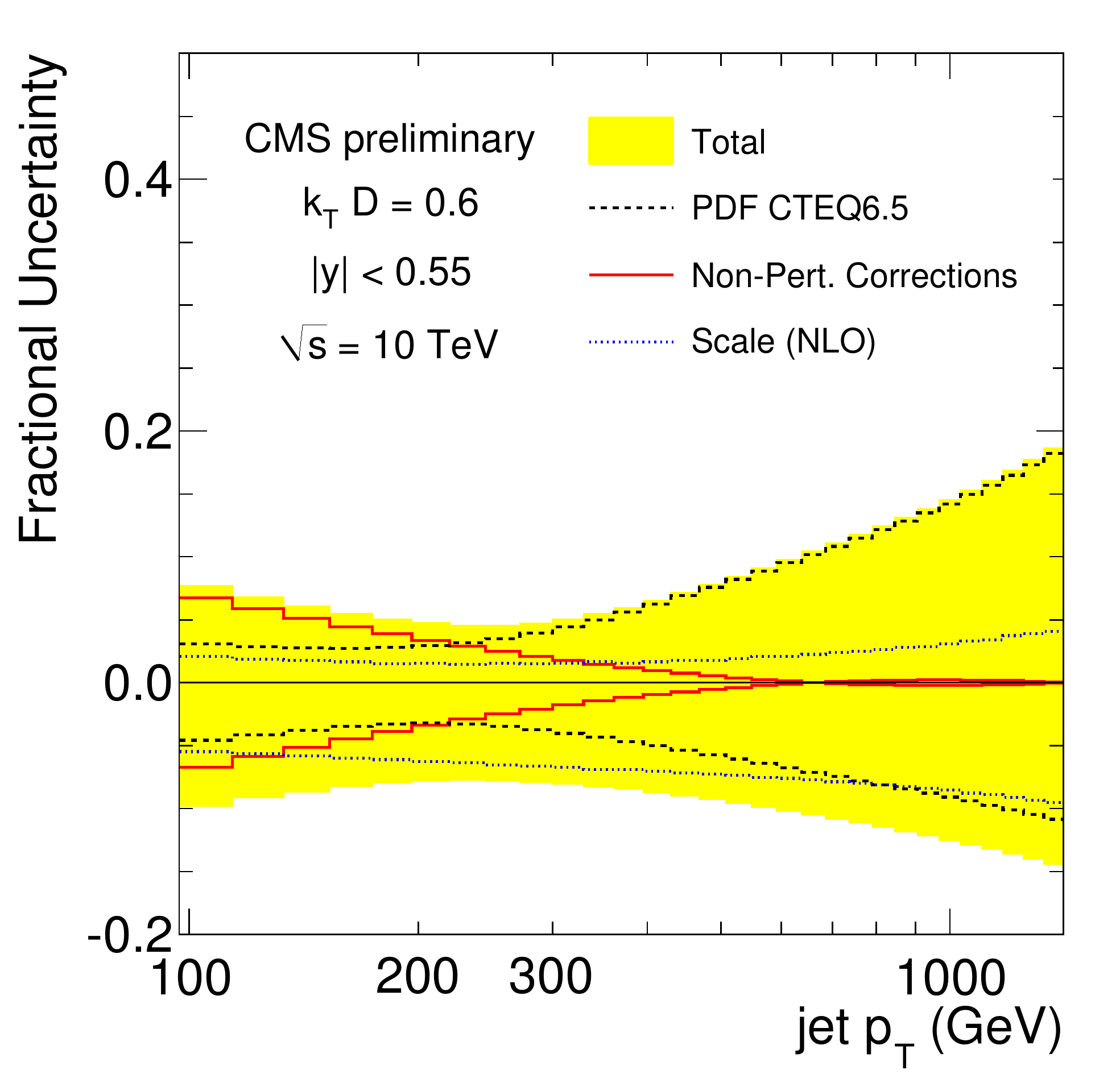}
  \caption{Fractional experimental (left) and theoretical (right)
    systematic uncertainties at central rapidity for the \kt jet
    algorithm. For better visibility the $y$-axis ranges have been
    chosen differently.}
  \label{fig:InclJetsSystUncertainty}
\end{figure}

The smeared \textit{Ansatz Function} which has been fitted to the
measured spectrum is also used to derive the uncertainties associated
with a flat (in \ptjet) $10\%$ jet energy scale uncertainty as well as
a $10\%$ variation relative to the nominal value of the JER as
estimated in~\cite{CMS-PAS-JME-09-007}. The result including an
assumed initial $10\%$ uncertainty on the luminosity is shown in
figure~\ref{fig:InclJetsSystUncertainty} left together with a summary
of the associated theory uncertainties on the right. The latter have
been evaluated using \NLOJETPP~\cite{Nagy:2001fj} and
\FASTNLO~\cite{Kluge:2006xs} for the PDF (CTEQ6.5~\cite{Tung:2006tb})
as well as scale uncertainties and the difference between
\PYTHIA~\cite{Sjostrand:2006za} and \HERWIGPP~\cite{Bahr:2008pv} for
the non-perturbative corrections.

Despite rather large experimental uncertainties initially, some
signals of new physics like contact interactions would be observable
already at start-up in jet cross sections at transverse momenta beyond
Tevatron energies. This is demonstrated in
figure~\ref{fig:InclJetsComparison} right where a contact interaction
term corresponding to a compositeness scale of $\Lambda^+ = 3\TeV$ is
drawn in addition to a pure \PYTHIA QCD prediction.

\subsection{Dijets}

\begin{figure}[htbp]
  \centering
  \includegraphics[width=0.50\textwidth]{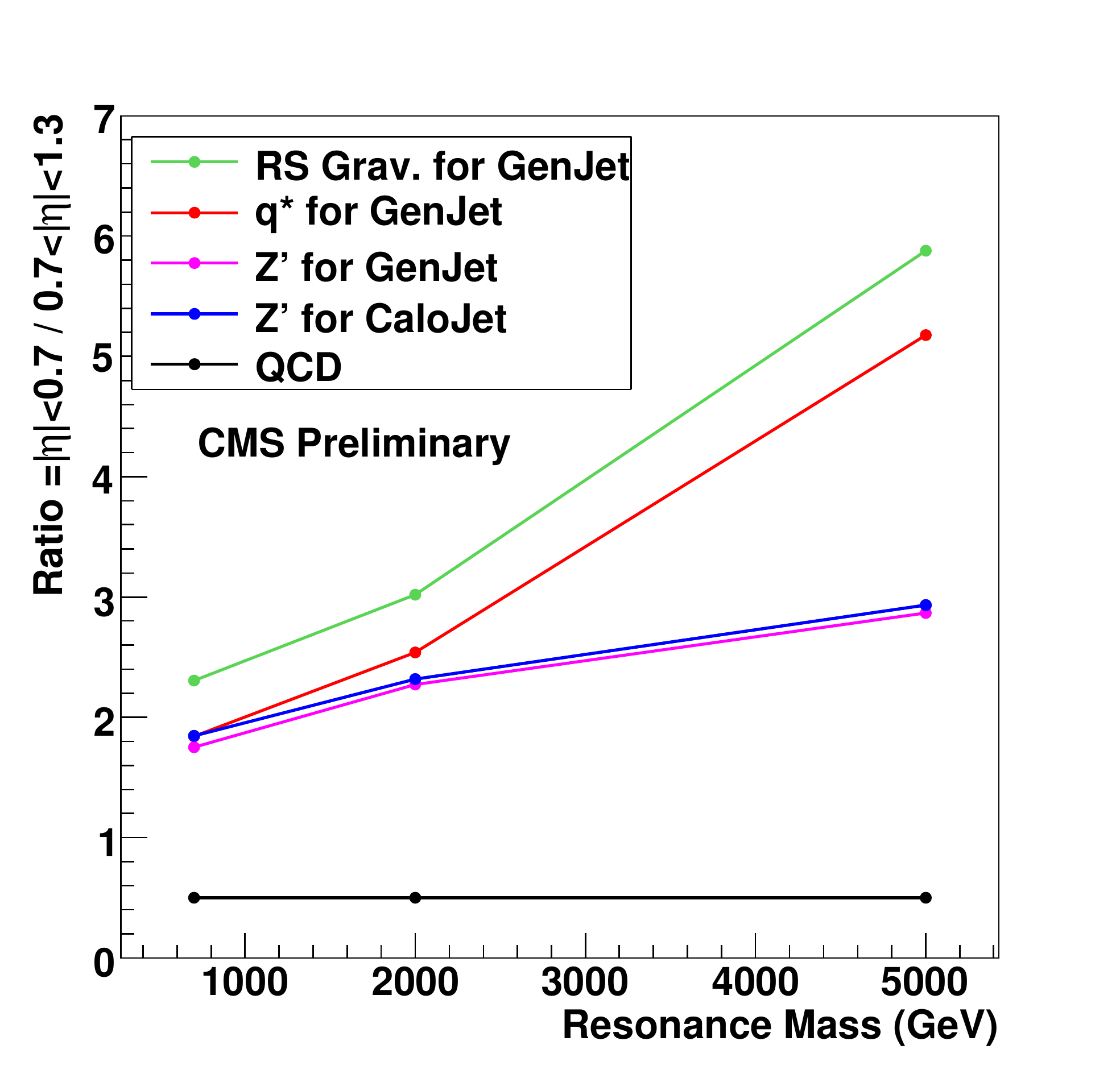}%
  \includegraphics[width=0.50\textwidth]{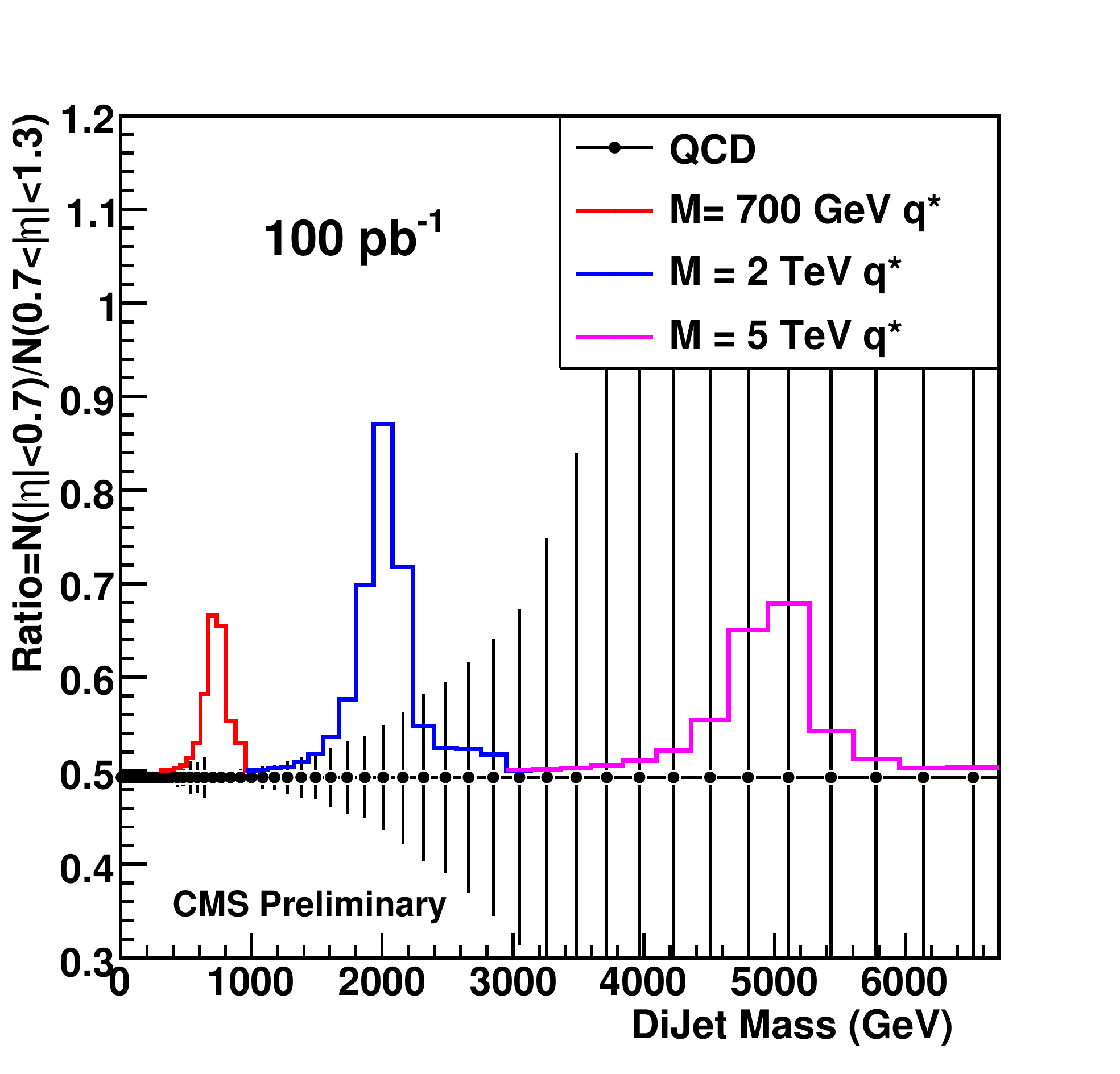}
  \caption{Dijet cross-section ratios in pseudorapidity versus
    resonance mass for \mbox{spin-2} Randall-Sundrum gravitons,
    \mbox{spin-1} $Z'$ bosons, \mbox{spin-1/2} excited quarks $q^*$
    and QCD (left). Dijet ratio versus resonance mass for three
    different excited quark masses compared to QCD (right) with
    statistical uncertainties as expected for $100\pbinv$ of
    integrated luminosity at $\sqrt{s}=14\TeV$.}
  \label{fig:DijetRatio}
\end{figure}

Another possibility to search for new phenomena already at start-up
with about $100\pbinv$ of integrated luminosity is to look for
resonances in the dijet mass spectrum e.g.\ from the decay of
\mbox{spin-2} Randall-Sundrum gravitons, \mbox{spin-1} $Z'$ bosons or
\mbox{spin-1/2} excited quarks $q^*$ as presented
in~\cite{CMS-PAS-SBM-07-001} for $\sqrt{s}=14\TeV$. Since these
resonances exhibit a more isotropic decay angular distribution than
dijets from QCD, it is possible to reduce or eliminate the sensitivity
to the dominant sources of experimental uncertainty from the JEC
respectively the luminosity determination by examining only the ratio
of the cross section in two different regions in
pseudorapidity. Figure~\ref{fig:DijetRatio} left shows the resulting
ratios of $\sigma_\text{dijet}(|\eta_j|<0.7)$ to
$\sigma_\text{dijet}(0.7<|\eta_j|<1.3)$ for three different resonance
masses which come out to be significantly larger than for QCD\@.
Figure~\ref{fig:DijetRatio} right illustrates for three different
masses of a potential $q^*$ resonance the observable dijet mass
spectrum in comparison to QCD including statistical uncertainties as
expected for $100\pbinv$.

\subsection{Dijet azimuthal Decorrelations}

\begin{figure}[htbp]
  \centering
  \includegraphics[width=0.50\textwidth]{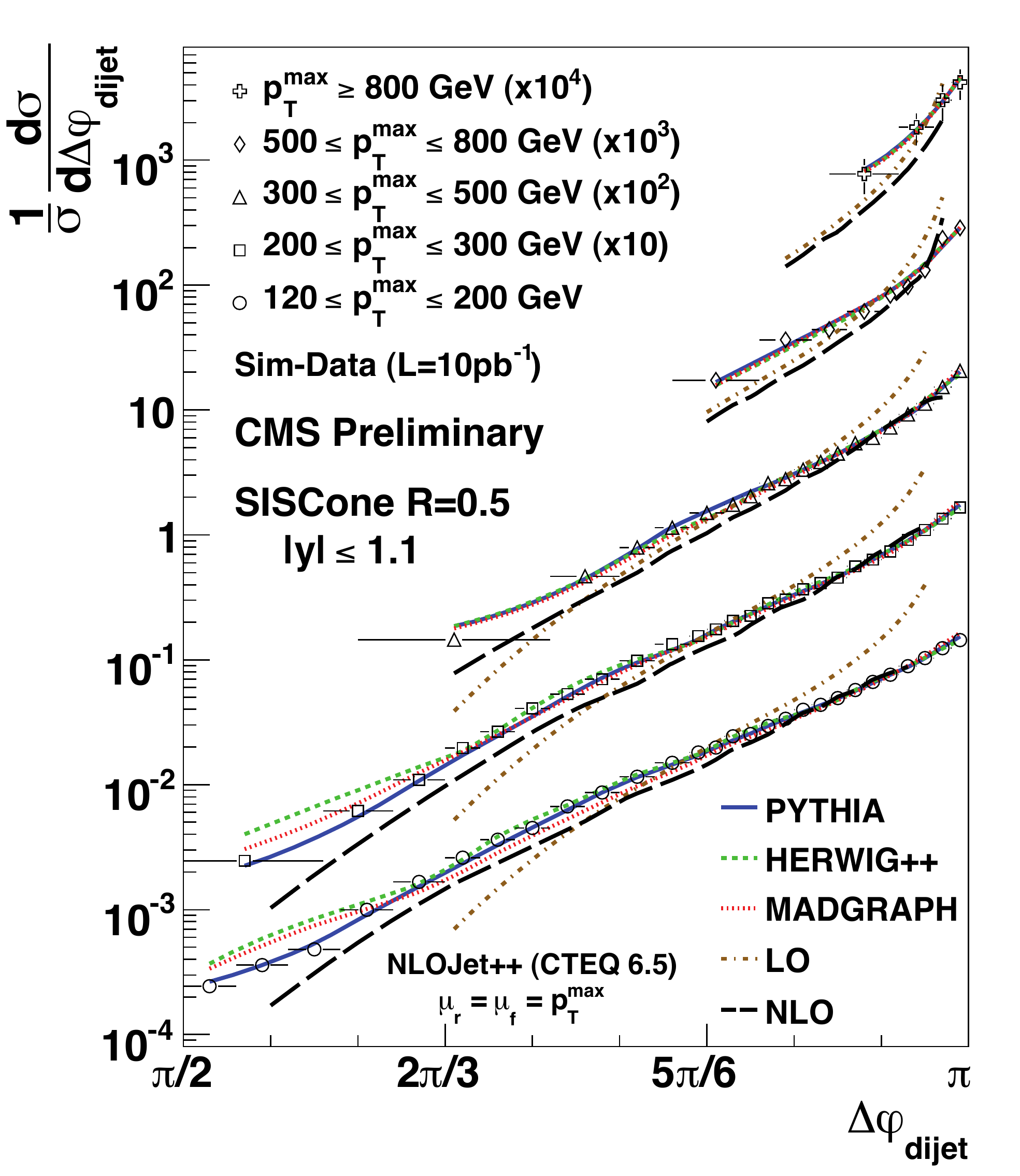}
  \caption{Corrected $\Delta \varphi_\text{dijet}$ distributions
    reconstructed from simulated energy depositions in the
    calorimeters (black symbols) are presented in several bins of
    leading jet \pt together with the statistical uncertainties as
    expected for $10\pbinv$ of integrated luminosity at $\sqrt{s} =
    10\TeV$. In addition, the distributions are compared to the
    particle jet predictions from \PYTHIA (full), \HERWIGPP
    (short-dashed), \MADGRAPH (dotted), and the predictions from LO
    (dash-dotted) and NLO pQCD (long-dashed line).}
  \label{fig:AzDecorr}
\end{figure}

In the study~\cite{CMS-PAS-QCD-09-003} of the normalized dijet cross
section $\frac{1}{\sigma_\text{dijet}} \cdot \frac{d
  \sigma_\text{dijet}}{d \Delta \varphi_\text{dijet}}$ emphasis is put
on the angular correlation in azimuth between the two leading jets
reconstructed from simulated energy depositions in the
calorimeters. Angular quantities in general can be measured more
precisely than the energy of jets as here the JEC uncertainty only
affects the classification of events into different bins of the
leading jet \pt.  The remaining total systematic uncertainty including
effects of the JER and the required unsmearing using different MC
generators is estimated to vary approximately linearly from $5\%$ at
$\Delta\varphi_{\text{dijet}} = \pi$ to $10\%$ at
$\Delta\varphi_{\text{dijet}} = 5\pi/9$.  In Figure~\ref{fig:AzDecorr}
the corrected $\Delta\varphi_\text{dijet}$ distributions from
simulated \PYTHIA events are compared in several bins of leading jet
\pt with the predictions of several MC generators and with LO as well
as NLO pQCD\@.

\subsection{Event Shapes}

\begin{figure}[htbp]
  \centering
  \includegraphics[width=0.50\textwidth]{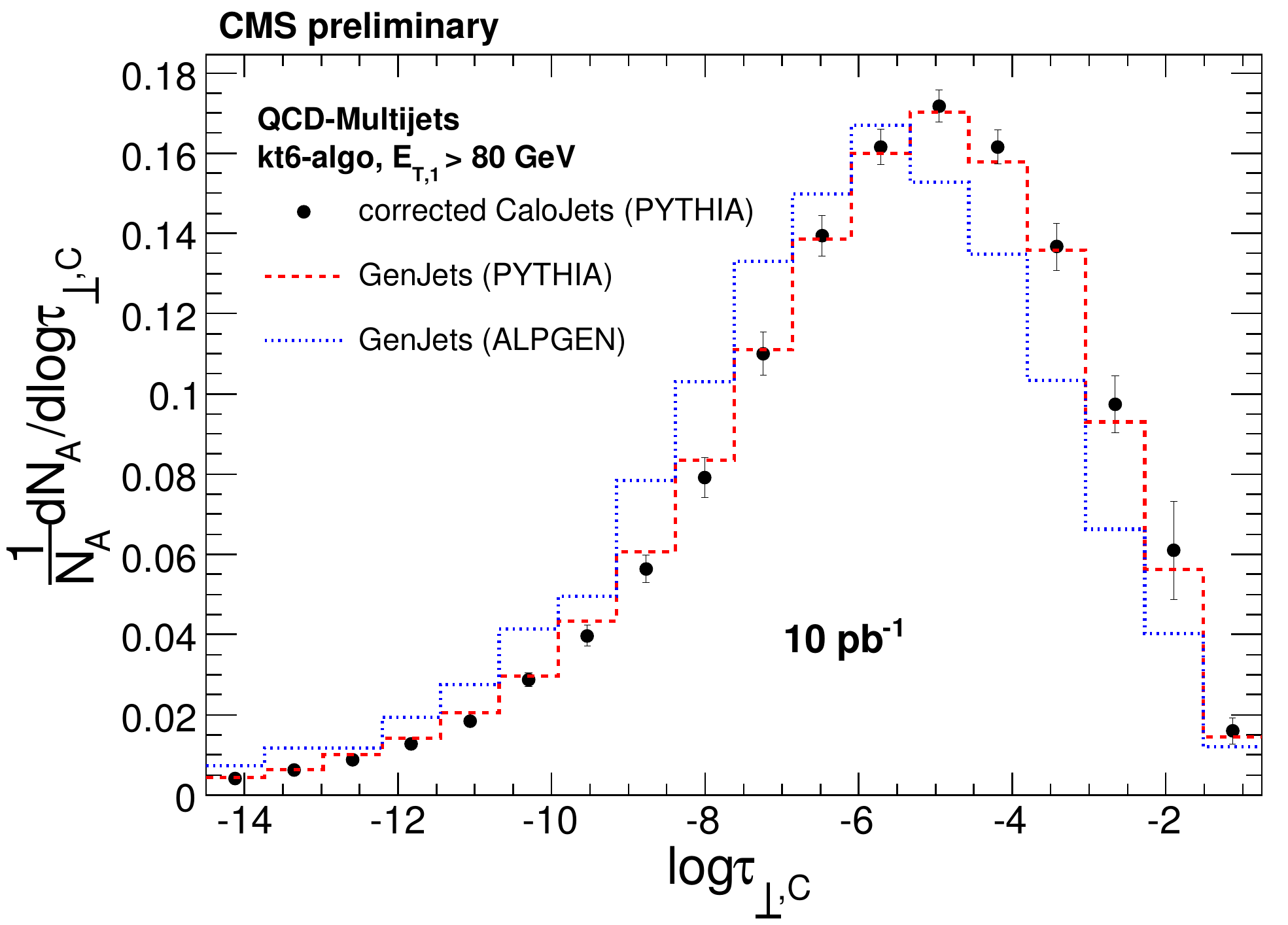}%
  \includegraphics[width=0.50\textwidth]{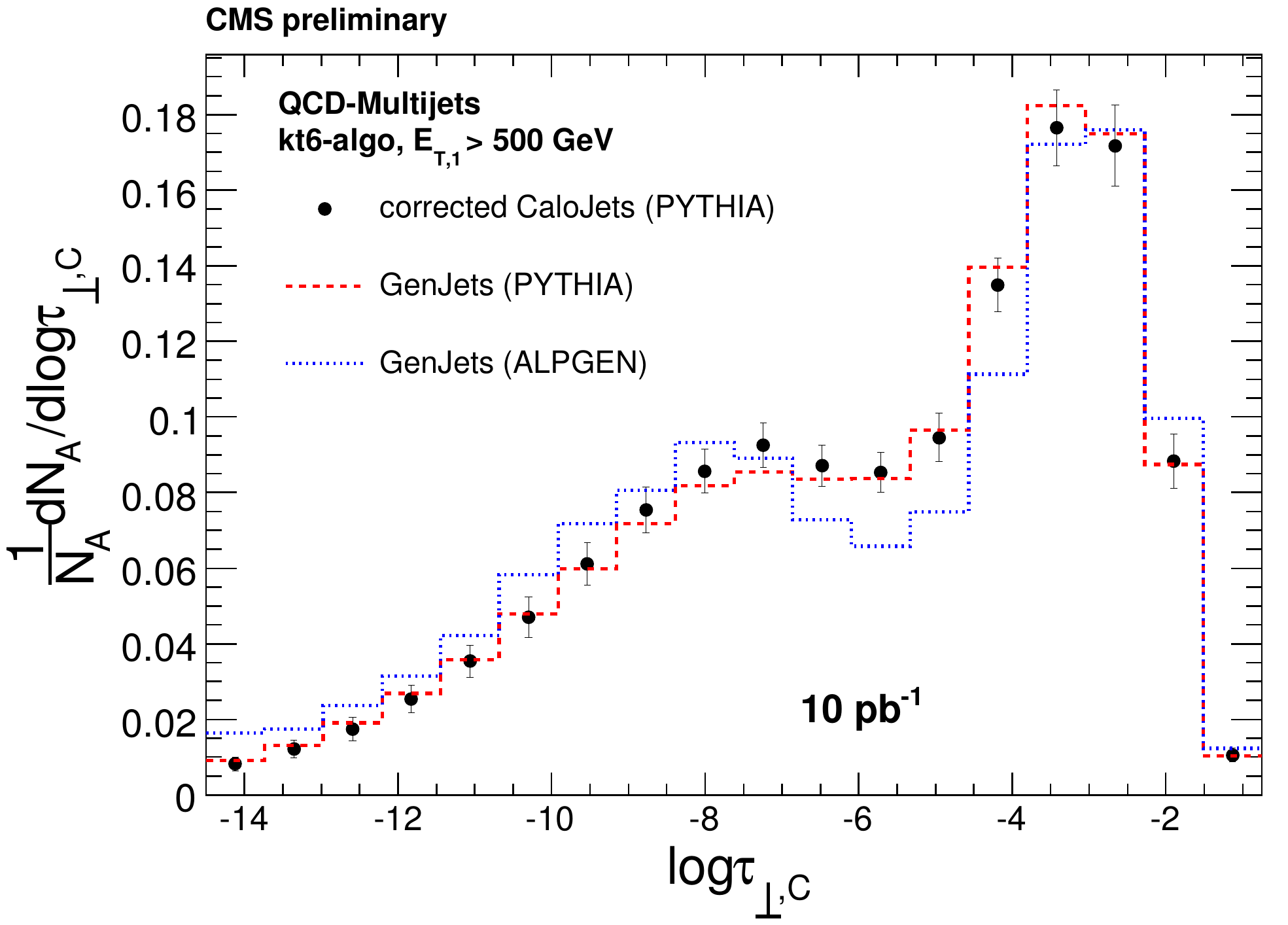}
  \caption{The central transverse thrust distribution
    ($\tau_{\perp,\mathbf{\mathcal{C}}}$, in logarithmic scale)
    reconstructed from simulated energy depositions in the
    calorimeters (black points) is presented for $E_{\text{T},1}
    ^{\text{cor}} > 80\, \text{GeV}$ (left) and $E_{\text{T},1}
    ^{\text{cor}} > 500\, \text{GeV}$ (right) together with the
    statistical and dominant systematic uncertainties as expected for
    $10\pbinv$ of integrated luminosity at $\sqrt{s} = 14\TeV$. A
    trigger pre-scale of $100$ is assumed in the left plot. In
    addition, the distributions are compared to the generator
    predictions of \PYTHIA (dashed) and \ALPGEN (dotted line).}
  \label{fig:EventShapes}
\end{figure}

Normalized hadronic event shape distributions of e.g.\ central
transverse thrust
$\tau_{\perp,\mathbf{\mathcal{C}}}$~\cite{Banfi:2004nk} which have
been analyzed in~\cite{CMS-PAS-QCD-08-003} are somewhat similar to the
previous observable in the sense that they characterize the geometric
momentum flow within an event. Only they usually exploit the complete
four-vectors of the measured objects. The sensitivity to the JEC is
then reduced by normalizing the quantity derived from the four-vectors
to the momentum sum. In general, event shapes, which have also been
measured in $e^+e^-$ and $ep$ collisions, do not necessarily require
the use of jet algorithms. In this study, however, the shape defining
objects have been chosen to be jets reconstructed from simulated
energy depositions in the calorimeters with pseudorapidities up to
$|\eta|< 1.3$. % Different jet algorithms have been employed.
Figure~\ref{fig:EventShapes} shows the
$\tau_{\perp,\mathbf{\mathcal{C}}}$ distribution for \kt jets with jet
size $D=0.6$ including statistical and systematic uncertainties from
JEC and JER for assumed $10\pbinv$ of integrated luminosity at
$\sqrt{s}=14\TeV$ together with the MC predictions from \PYTHIA and
\ALPGEN. With higher transverse momenta more events approach thrust
values corresponding to a dijet configuration
($\tau_{\perp,\mathbf{\mathcal{C}}} \rightarrow 0$) as can be seen
from a comparison of the distributions for two different minimal
transverse energies of the leading jet in figure~\ref{fig:EventShapes}
left and right. Early measurements of event shapes allow to study
differences in the modelling of QCD multi-jet production and are a
valuable input to MC generator tuning.

\subsection{Jet Shapes}

\begin{figure}[htbp]
  \centering
  \begin{minipage}[t]{0.48\textwidth}
    \centering
    \includegraphics[width=\textwidth]{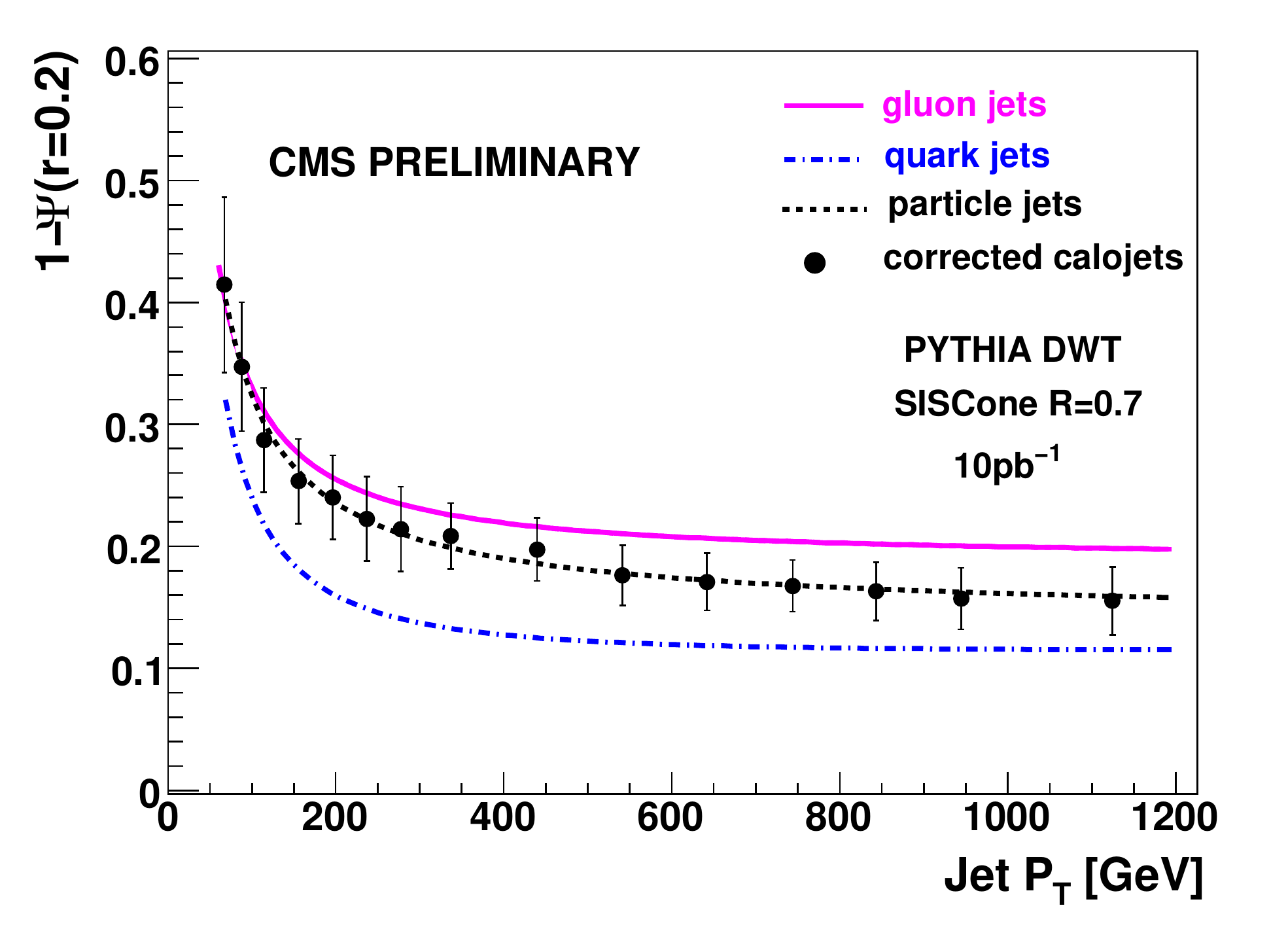}
    \caption{The fractional transverse momentum of a jet outside
      $R$=0.2, $1- \psi(0.2)$, is presented versus \ptjet for jets
      reconstructed from simulated energy depositions in the
      calorimeters (black points) in the rapidity region $|y|<1$
      including statistical and systematic uncertainties as expected
      for $10\pbinv$ of integrated luminosity at $\sqrt{s} =
      14\TeV$. In addition, \PYTHIA (tune DWT) predictions are shown
      for quark initiated (dash-dotted), gluon initiated (solid), and
      for all particle jets (dashed line).}
    \label{fig:JetShapes14}
  \end{minipage}
  \hspace{0.02\textwidth}
  \begin{minipage}[t]{0.48\textwidth}
    \centering
    \includegraphics[angle=90,width=\textwidth]{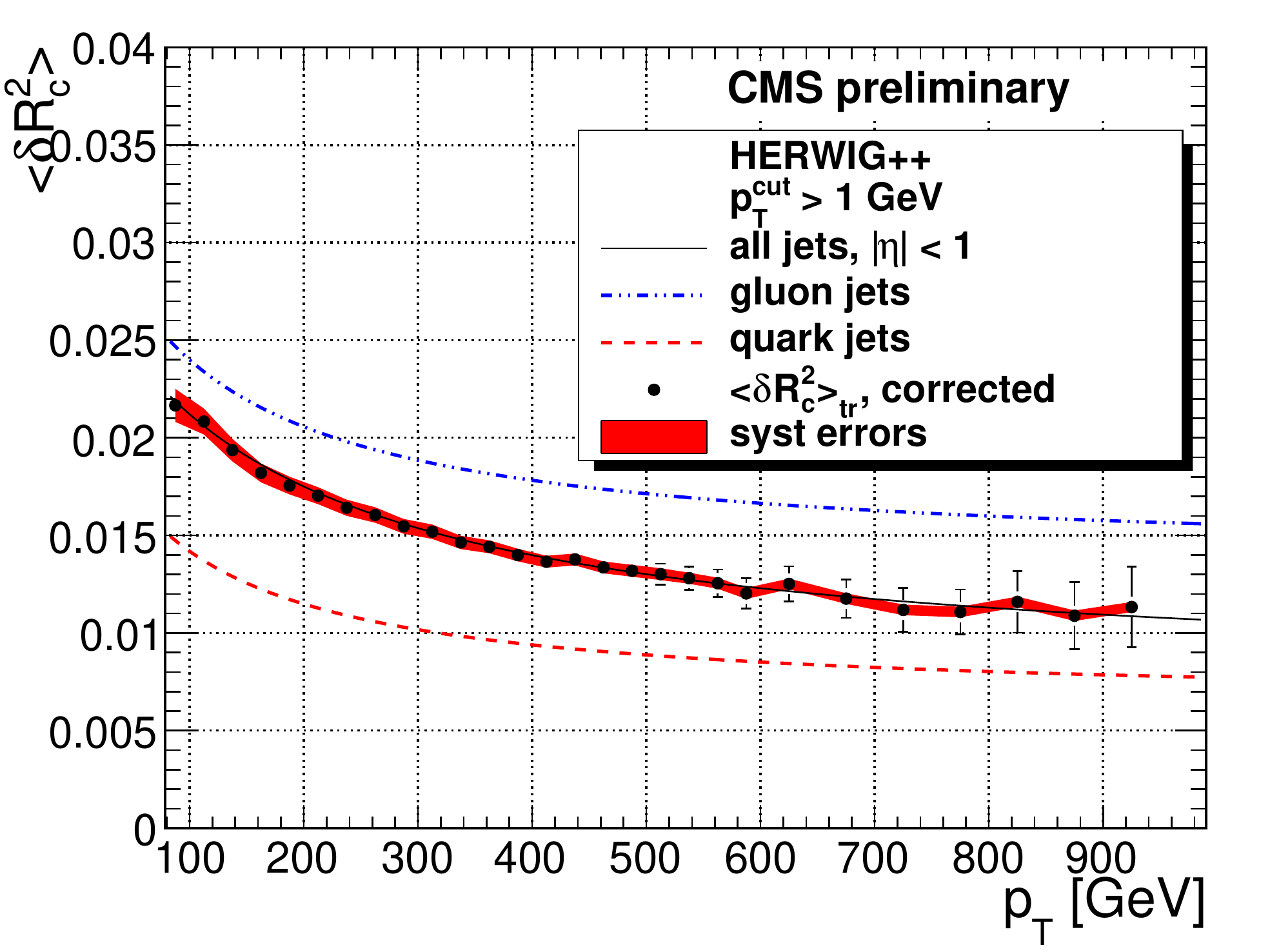}
    \caption{The second central moment $\left<\delta R^{2}_{c}\right>$
      of the jet transverse energy distribution is presented versus
      \ptjet for jets reconstructed from tracks (black points) in the
      pseudorapidity region $|\eta| < 1$ including statistical
      uncertainties as expected for $10\pbinv$ of integrated
      luminosity at $\sqrt{s} = 10\TeV$.  The total systematic
      uncertainty is indicated by the shaded region.  In addition,
      \HERWIGPP predictions using charged particles are shown for
      quark (dashed) and gluon initiated jets (dash-dotted line).}
    \label{fig:JetShapes10}
  \end{minipage}
\end{figure}

As the last topic to be covered in this note
jet shapes look into the internal structure of jets.
Two observables are suggested in~\cite{CMS-PAS-QCD-08-005}
and~\cite{CMS-PAS-QCD-08-002}: The fractional transverse momentum
$1-\psi(R)$ of a jet outside the jet core with a radius of $R=0.2$ and
the second central moment of the jet transverse profile $\left<\delta
  R^{2}_{c}\right>$. To explore the largest jet \pt range possible,
the first quantity has been evaluated for the two leading calorimeter
jets from QCD dijet production and is compared in
figure~\ref{fig:JetShapes14} with the MC prediction of \PYTHIA for
quark and gluon jets. The total uncertainty is dominated at low \pt by
systematic uncertainties due to the JEC, non-linearities of the
calorimeter and fragmentation model dependencies estimated using
\PYTHIA and \HERWIGPP. At high \pt the uncertainty due to lack of
statistics with only $10\pbinv$ at $\sqrt{s}=14\TeV$ takes over.

To reduce the sensitivity to the JEC and non-linearities in the
calorimeters the second central moment $\left<\delta R^{2}_{c}\right>$
has been calculated for jets where the tracks of charged particles
($\pt>1\GeV$) associated with a jet are used to correct the
calorimetric energy determination~\cite{CMS-PAS-JME-09-002}. The jet
substructure then is derived from tracks respectively the charged
particles alone. Instead of $10\%$ JEC uncertainty in the former study
only $\approx 5\%$ are assumed in the
latter. Figure~\ref{fig:JetShapes10} compares the result for
$\left<\delta R^{2}_{c}\right>$
% , including error bars for the statistical and an error band for the
% systematic uncertainties,
with the MC prediction of \HERWIGPP for quark and gluon jets.

Both observables will serve as input for tuning MC generators, in
particular with respect to fragmentation models, and may allow an
extraction of the quark-gluon jet fraction.

\section{Outlook}

A number of QCD analyses involving charged particle tracks and jets
possible already after a few days of data taking or after accumulating
roughly $10$ to $100\pbinv$ of integrated luminosity have been
presented. A rich program of new physics measurements re-establishing
the Standard Model and preparing the arena for searches for new
phenomena will be possible.  CMS is well prepared and first
proton-proton collisions at the LHC are eagerly awaited.

\section{Acknowledgements}

I would like to thank the conference organizers for the kind
invitation and for the overwhelming hospitality I had the occasion to
encounter.  I would also like to thank Kirsti Aspola for her kind
organizational help.

\newpage
%
% BibTeX Bibliography
%
\bibliographystyle{lucas_unsrt}
\bibliography{my-references,cms-pub,qcd}

\end{document}